\documentclass[12pt]{article}
\pdfoutput=1
\usepackage[body={180mm,250mm}]{geometry}
\usepackage{amsfonts,amsmath}
\newlength{\abstractwidth}
\usepackage{epsf}
\usepackage{a4wide}
\usepackage{epsfig}

\usepackage[margin=1cm]{caption}

\newcommand{\bea}{\begin{eqnarray}}
\newcommand{\eea}{\end{eqnarray}}
\def\no{\nonumber}
\newcommand{\half}{{1 \over 2}}

\newcommand{\p}{\partial}

\begin{document}
\setcounter{footnote}{0}

\title{{\bf Spin-2 spectrum of defect theories}}
\author{Constantin Bachas $^\flat$  and John Estes $^\natural$}

\maketitle

\centerline{$^\flat$ Laboratoire de Physique Th\'eorique de l'Ecole Normale Sup\'erieure, }
\centerline{24 rue Lhomond, 75231 Paris cedex, France}
\centerline{{\em bachas@lpt.ens.fr}
}

\vskip -9.5cm
\rightline{LPTENS-11/08}
\vskip 9.5cm

\vskip 3mm

\centerline{$^\natural$ Instituut voor Theoretische Fysica, Katholieke Universiteit Leuven, }
\centerline{Celestijnenlaan 200D B-3001 Leuven, Belgium}
\centerline{{\em johnalondestes@gmail.com}
}


\abstract{ We study spin-2 excitations in the background of the recently-discovered type-IIB
solutions of D'Hoker et al. These are holographically-dual to defect conformal field theories,
and they are also of interest in the context of the Karch-Randall  proposal for a string-theory embedding
of localized gravity.  We first  generalize an argument by Csaki et al  to show that for any solution with
four-dimensional  anti-de Sitter,  Poincar\'e  or de Sitter invariance
 the spin-2 excitations obey the massless scalar wave equation in ten dimensions.
For the interface  solutions at hand this reduces to a Laplace-Beltrami
equation on a Riemann surface with disk topology, and in the simplest case of the supersymmetric Janus
solution it further reduces to an ordinary differential equation known as Heun's equation.
We solve this equation numerically,  and exhibit the spectrum as a function of the
dilaton-jump parameter $\Delta\phi$. In the limit of large $\Delta\phi$
a nearly-flat linear-dilaton dimension grows large, and the
 Janus geometry becomes effectively five-dimensional.  We also discuss the
difficulties of localizing four-dimensional gravity in the  more general
backgrounds with NS5-brane  or D5-brane charge, which will be
analyzed in detail  in a companion paper.
   }

\vfill\eject

\tableofcontents
\vfill\eject

\baselineskip=15pt
\setcounter{equation}{0}
\setcounter{footnote}{0}

\renewcommand{\theequation}{\arabic{section}.\arabic{equation}}

\section{Introduction}
\setcounter{equation}{0}

Whether gravity can be localized
\cite{RubS, Randall:1999vf, Dvali:2000hr} is a question of great interest for cosmology.
  It is closely related to the question of whether the graviton can have a mass \cite{vanDam:1970vg,Zak,
  Vain,Boulware:1973my},
    or more generally
  whether Einstein's theory -- with or without a cosmological constant  term --
    can be consistently modified  at cosmic-distance  scales.
   Despite many interesting results
  (see e.g. \cite{Gregory:2000jc, ArkaniHamed:2002sp, Deffayet:2005ys, Rubakov:2008nh,Kiritsis:2008at,Blas:2008uz,
  deRham:2010gu,Alberte:2010qb}
  and references therein),
   there is still no  definitive  answer  to these questions.

   The difficulties with   low-energy descriptions of  localized and/or  massive gravity typically
   arise  when one considers  classical non-linearities,  and quantum corrections.  Problems  include
   the non-decoupling of  heavy mass scales
   (the Planck scale, the bulk curvature or the brane thickness) and/or the appearance of ghosts.
  Clearly,  an embedding
   in an ultraviolet-complete theory like string theory could  shed important
   light on these issues.  A promising proposal for such an embedding
 was made some time ago by Karch and Randall  \cite{KR1, KR2}. These authors  considered
  an AdS$_4\times$S$^2$ brane in an AdS$_5\times$S$^5$ bulk, and argued that
  a nearly-massless graviton can be obtained by  tuning   the brane tension.
  In   their calculation  Karch and Randall had to approximate  the background geometry  by two AdS$_5$ slices
  glued together along a ``thin"  AdS$_4$ brane.  The exact type-IIB supergravity solutions, describing the full
  back-reaction of  branes on geometry,  were  discovered only recently by
    D'Hoker et al \cite{DEG1, DEG2}
    (see also \cite{Gomis:2006cu,Lunin:2008tf} for related work).
    A central  motivation for our  work here was to
  check  whether the conclusions of  Karch and Randall   survive
   in these  exact  string-theory backgrounds.

   The supergravity solutions of  D'Hoker et al  are holographically-dual to defect conformal field theories
     \cite{Karch:2000gx,DeWolfe:2001pq,Bachas:2001vj,Erdmenger:2002ex, Gaiotto:2008sd}, i.e. to two or more conformal theories
     interacting along a scale-invariant domain
     wall.\footnote{To be distinguished from  conformal theories
     interacting via  multi-trace marginal operators  in the bulk,  and  which have been argued to be the duals of multi-gravity models
  \cite{Kiritsis:2006hy, Aharony:2006hz, Kiritsis:2008at}. Their embedding in string theory is more problematic, as it   most likely involves
quantum-gravity effects.}
 The ``locally-localized graviton" has been
discussed from the viewpoint of the defect CFTs in  \cite{Aharony:2003qf}
(for earlier perspectives see \cite{Porrati:2001gx,Bousso:2001cf,Duff:2004wh},  and for a recent one
 \cite{Aharony:2010ay}
).
 Other than their potential relevance to the  localization of gravity,
our results are  thus also of  interest in this dual context.

    In this paper we will first set up the spectral problem for spin-2 excitations in the   interface geometries of
 refs. \hskip -0.7mm  \cite{DEG1, DEG2}. These  have the structure of an AdS$_4\times$S$^2\times$S$^2$ spacetime,
 fibered over a Riemann surface $(\Sigma, \hat g )$  with disk  topology. The excitations which are constant on  the
 spheres, and which have spin 2 in AdS$_4$,  obey Neumann boundary conditions on the surface $\Sigma$. Their
  shifted mass-squared operator,  $m^2 + 2$,   is, as we will see,  the Laplace-Beltrami  operator on the disk with
    metric  given by $\bar g = e^{-2A} \hat g$,   where $A$ is  the AdS$_4$  warping factor.
  In  deriving  this result,   we will generalize slightly an argument of  Csaki  et al \cite{Csaki:2000fc} to show
   that the spin-2  excitations obey the massless scalar-wave equation in ten dimensions.  The argument, first made
   in the context of  flat thick branes and  scalar fields,\footnote{For branes of co-dimension one,  the observation was actually made
   earlier  in ref. \cite{Brandhuber:1999hb}.}
  generalizes easily to all supergravity solutions  with 4D anti-de Sitter, Poincar\'e or de Sitter symmetry,  and
  with arbitrary $p$-form flux  backgrounds.
 The universality of the spin-2  wave equation should be contrasted with
 what happens for spins $\leq 1$, whose wave equations
 depend  a priori on non-metric features  of the backgrounds.

Solving the spectral problem on the surface $(\Sigma, \bar g)$  is, in general,  a difficult  task. It simplifies considerably
 for the supersymmetric Janus solution [which describes a supersymmetric dilaton domain wall], for which the
 relevant eigenmode equation is separable.  The reduced
 ordinary differential equation has four regular singular points,
 and  is known  in the mathematics  literature as   Heun's equation \cite{heun}. We will solve this equation numerically,
 and exhibit the spectrum as a function of the dilaton-jump parameter $\Delta\phi$.
 As we will see, Janus cannot localize four-dimensional gravity  and its only interesting  limit,
  $\Delta\phi \to \infty$,   is a limit  in which a flat fifth dimension  decompactifies.
This behavior can be attributed to the vanishing superpotential of the dilaton field, whose domain walls have
 no  intrinsic tension or  thickness. Solutions with  NS5-brane  and/or D5-brane charge are in this respect
 more promising, in line with the original proposal of Karch and Randall \cite{KR2}.
 Generating a large scale hierarchy without decompactifying extra dimensions looks, however, a priori
 a difficult task.
 The  detailed analysis of the backgrounds with five-brane charge, for which the eigenmode equation on the strip
 does not factorize,  will be presented
  in a forthcoming companion publication \cite{tocome}.


 This paper is organized as follows: in section 2 we derive the linearized equation for transverse-traceless graviton
 modes in any geometry with AdS,  dS  or  Poincar\' e invariance in four dimensions. Generalizing the argument
 of  \cite{Csaki:2000fc}  we show that,  in all cases,
  these modes obey the massless scalar wave equation in ten dimensions.
   In section 3 we review the solutions  of D'Hoker et al \cite{DEG1, DEG2}, and relate  their free parameters
   to  brane charges and to the asymptotic values of the dilaton field.
  Section 4   reviews  the qualitative features  responsible for the
  localization of the graviton in the  protoype thin-brane geometry of Karch and Randall \cite{KR1}.
  We show that the Janus solution does not have the required  ``warp-factor bump",  and we describe the
  large-$\Delta\phi$ limit of its  geometry. We also discuss briefly the
  backgrounds with five-brane charge, whose detailed analysis will be presented in ref. \cite{tocome}.
  In section 5 we reduce the eigenmode equation for all interface solutions
     to a spectral problem on the two-dimensional strip $\Sigma$, paying  special attention to
    the boundary conditions and  the norm of  the modes.  We verify our  formulae  by  rederiving  the well-known AdS$_5\times$S$^5$ spectrum.
  Finally, section 6 presents our analysis of the Janus spectrum.
  We reduce the  equation on $\Sigma$   to Heun's  equation \cite{heun}, and solve this latter numerically.
  The results are not surprising, and agree with expectations  in the large-$\Delta\phi$ limit. Some conventions
  on type-IIB supergravity are collected in the appendix A.

\section{Universal equation for spin-2 excitations}
\label{sec2}
\setcounter{equation}{0}

In this section we will derive  the equation for the spectrum of the spin-2 Kaluza-Klein excitations
around any solution  with maximal symmetry in four dimensions.  As we will see, the relevant second-order operator
is the massless scalar wave operator in ten (or in eleven)  dimensions.
It is an interesting fact that this mass operator
 only depends on the background metric, and on no other fields. The reason is that
   all other  background  fields
 enter the linearized equations  for   spin-2 modes  via the total energy-momentum tensor,  which can be
 re-expressed in terms of the background curvature.
This was  first observed  by Csaki et al  \cite{Csaki:2000fc} for   flat thick branes
 generated by  scalar fields.  We  here extend the argument   to  all supergravity solutions
with maximal (AdS,  dS  or Poincar\'e) symmetry, and with arbitrary $p$-form flux backgrounds.

\subsection{Ansatz for the Kaluza-Klein spin-2 modes}

   We are interested in  solutions of the supergravity equations  with a background metric that has the following warped form:
 \bea\label{ansatz}
 \widehat{ds^2} =    e^{2A(y)}  \bar g_{\mu\nu}(x)\, dx^\mu dx^\nu  +   \hat g_{ab}(y)\,  dy^a dy^b  \ .
   \eea
   Here  Greek indices run over $\{ 0,1,2,3\} $,
   $\bar g_{\mu\nu}$ is the metric for the unit-radius constant-curvature spacetime
  ($ \bar{\cal M}_4=  {\rm AdS}_4,  \mathbb{M}_4$ or   ${\rm dS}_4$,   for $k=-1, 0, 1$), and
   Latin indices  label the remaining coordinates, $y^a$, of the ``internal"  space $\hat{\cal M}_{d-4}$.
    The internal space may,  but need not,  be compact. In this section the total spacetime dimension is arbitrary,
   later we will focus on the type IIB supergravity for which  $d=10$.

We   restrict attention to  perturbations of the four-dimensional metric:
\bea\label{ansatz1}
d {s}^2  =  e^{2A} \left( \bar g_{\mu\nu} + h_{\mu\nu} \right)  dx^\mu dx^\nu
+ \hat g_{ab}\,  dy^a dy^b\ ,
\eea
and look for factorized solutions of the linearized field equations of the following form:
\bea\label{ansatz2}
h_{\mu\nu}(x, y) \, =\,  h_{\mu\nu}^{\rm [tt]} (x\vert  \lambda)\, \psi(y\vert \lambda)\  ,
\eea
 where $h_{\mu\nu}^{\rm [tt]}$ solves the Pauli-Fierz equations \cite{fierz} for  a massive spin-2 particle  in $\bar{\cal M}_4$.
 The Pauli-Fierz  equations read (see for instance \cite{Buchbinder:1999ar, Deser:2001wx})
 \bea\label{massivespin2}
(\bar  \Box_{x}^{\rm (2)} -  \lambda)\,  h^{\rm [tt]} _{\mu\nu}  = 0\qquad {\rm and}\qquad
\bar \nabla^\mu h_{\mu\nu}^{\rm [tt]} = \bar g^{\mu\nu}  h_{\mu \nu}^{\rm [tt]} = 0\  ,
 \eea
where $\bar\Box_{x}^{\rm (2)}$      is the Laplace operator acting on two-index tensors in $\bar {\cal M}_4$.
The last two equations, which  force $h_{\mu \nu}^{\rm [tt]}$ to be transverse and traceless (henceforth ``tt")  in four dimensions,
are non-dynamical
phase-space constraints.
As we will show, with the above ansatz the linearized Einstein equations reduce to a single second-order equation for the internal-space
wavefunction $\psi(y)$.  Normalizable solutions correspond to the allowed  eigenvalues of the wave operator, $\lambda$.
 The four-dimensional  Pauli-Fierz mass  is  related to $\lambda$ as  $m^2 =  \lambda- 2k$  \cite{Deser:2001wx}.

\vskip 1mm

The last two
equations in  (\ref{massivespin2}) eliminate  five out of the ten  components of  the
symmetric two-index tensor. The remaining degrees of freedom are what one needs to describe the
states of  a massive spin-2 particle in four dimensions. We note that the above metric perturbations are   transverse-traceless
 also  in $d$ dimensions,
\bea\label{TT}
\hat\nabla^M\,  \delta g_{MN} = \hat g^{MN}\,  \delta g_{MN} = 0
\  ,
\eea
where here  the   covariant derivatives   refer   to  the (bloc-diagonal)  metric
$\hat g_{MN} = (e^{2 A } \bar g_{\mu\nu} ,  \hat g_{ab} )$,  and
$\delta g_{MN} = (e^{2 A} h_{\mu\nu} ,  0 )$ is the  unscaled metric  perturbation.
 To verify  this statement  one  uses  the fact
 that  the  components $\hat\Gamma_{\mu\nu}^\rho$ of the affine connection are identical to those on $\bar {\cal M}_4$.
The converse of this statement is however false, i.e. the $d$-dimensional transverse-traceless conditions
 do not  imply the four-dimensional ones.
Thus our ansatz does not describe all metric perturbations,
but only those which have spin 2 in four dimensions.


\subsection{Reduction of the Einstein equations}

Turning now  to  the  Einstein equations, we  want to compute the linearized variation of the
Einstein tensor and of  its source term, i.e. the energy-momentum tensor of the matter fields.
Let us start with the Einstein tensor.
The variations of the affine connection and of the
curvature  are given,  at linear order, by the  expressions  \cite{Carroll:2004st}:
\bea
&&\delta \Gamma^K_{MN}
 =  {1\over 2}\hat g^{K\Sigma}\,  \left(  \hat\nabla_M \delta g_{N\Sigma} +
  \hat\nabla_N \delta g_{M\Sigma}  - \hat\nabla_\Sigma \delta g_{MN}\right)  \ , \nonumber  \\
 \  &&  \nonumber\\
 &&\delta R^{P}_{\  M K N} =  \hat\nabla_K (\delta \Gamma^P_{NM})  -  \hat\nabla_N (\delta \Gamma^P_{K M}) \ .
\eea
Using the [tt] property  of  $\delta g_{MN}$  leads, after a little algebra,  to the following linear variation of the Ricci tensor:
\bea\label{linRicci}
\delta R_{MN} \hskip -2mm
&=&   {1\over 2} [\hat\nabla_K, \hat\nabla_M]  \delta g_N^{\ K}
 +  {1\over 2} [\hat\nabla_K, \hat\nabla_N]  \delta g_M^{\ K} - {1\over 2} \hat\nabla_K \hat\nabla^K \delta g_{MN}  \nonumber \\
 &=& {1\over 2} \left(
 \hat R_{PM}\,  \delta g_N^{\ P} -  \hat R^P_{\ NKM}\,  \delta g_P^{\ K} +  (N\leftrightarrow M)
 \right)  -  {1\over 2}\, \hat\nabla^K \hat\nabla_K  \delta g_{MN}  \, ,
\eea
where indices are raised with the  background metric  $\hat g_{MN}$.
 It is at this stage convenient to  bring the metric by
a  Weyl rescaling to  direct-product form [we use bars for the rescaled background,  as opposed to hats for the original one]:
\bea
{\bar g}_{MN}  \equiv e^{-2 A} \hat g_{MN}
=  (\bar g_{\mu\nu} ,  e^{-2 A } \hat g_{ab} )\ .
\eea
The  curvature and Ricci tensors   in (\ref{linRicci}) can be expressed
in terms of  the corresponding ``barred" tensors, and derivatives of the conformal factor
 (see e.g. \cite{{Dabrowski:2008kx}} for a collection of Weyl-transformation formulae). The terms relevant for our purposes
  here are:
\bea\label{ids}
&&\hat R^P_{\ NKM} =  \bar R^P_{\ NKM} +
(\bar g^P_{\ M}\, \bar g_{NK} - \bar g^P_{\ K}\, \bar g_{NM})  \,  A_{; \Sigma}  \,  A^{; \Sigma} \,    + \cdots\ ,  \nonumber  \\
\ &&  \nonumber   \\
&& \hat R_{PM} =  \bar  R_{PM} -  \bar g_{PM} \left[  \, \bar\Box_y\, A  +
 (d-2)  \, A_{; \Sigma}  \,  A^{; \Sigma} \,
 \right]  + \cdots\ ,
\eea
 where $d$ is the total spacetime dimension, $\bar\Box_y$ is the scalar Laplacian in the internal space with metric $\bar g_{ab} =
e^{-2 A} \hat g_{ab}  $, and
$A_{; \Sigma}  \equiv \partial_\Sigma A$.
The dots stand for  omitted terms  that
vanish when inserted in  ({\ref{linRicci}),  either because  the perturbation
 $\delta g_{MN}$ is  traceless,   or because it points  in the $\bar {\cal M}_4$ directions
 (so that  its contraction with $A_{;M}$ is zero).
  Note that the second
  line in (\ref{ids}) is not just the contraction
 of the first line,  because the terms  omitted in these two lines are different.

 The Weyl transformation of the third term  in expression (\ref{linRicci}) requires   some more work.
 From the  transformation property
 of the affine connection,
 \bea
 \hat \Gamma^{K}_{MN} = \bar \Gamma^{K}_{MN} + (\bar g^K_{\ M} \, A_{;N} +  \bar g^K_{\ N} \, A_{;M}  - \bar g_{MN} \, A^{;K} )\ ,
 \eea
and the fact that in our case $A^{;\Sigma}\, \delta g_{\,\Sigma \Xi} = 0$ one finds first:
\bea
\hat\nabla_K \,   \delta g_{MN} &=& \bar\nabla_K\,   \delta g_{MN} - 2A_{;K}\delta g_{MN} - A_{;M}\delta g_{KN} - A_{;N}\delta g_{MK}
\nonumber \\ \ && \nonumber\\
&=& e^{2A}\, \left(  \bar\nabla_K\,   h_{MN}- A_{;M} h_{KN} - A_{;N} h_{MK} \right) \ ,
\eea
where  $h_{MN} \equiv e^{-2A} \delta g_{MN}$ is the rescaled perturbation.
  Applying one more derivative, and using
again the special properties of $h_{MN}$,  gives after some straightforward algebra,
 \bea\label{boxT}
\hat\nabla^K \hat\nabla_K \,   \delta g_{MN} =
  \bar\nabla^K \bar\nabla_K\,  h_{MN}  + (d-2) \, A_{;K}  \bar\nabla^K\,
    h_{MN}  -  2  \, A_{; \Sigma}  \,  A^{; \Sigma} \, h_{MN}  \ .
\eea

 \vskip 1mm

Inserting  now the expressions
 (\ref{ids}) and (\ref{boxT})  in the second line of ({\ref{linRicci}) gives
 the variation of the Ricci tensor, expressed  in terms of the  rescaled
metric ${\bar g}_{MN}$, its curvature tensor,  and  the warp  factor $A(y)$.
The advantage of this rewriting is that
 ${\bar g}_{MN}$ is a direct-product metric, so that   its
  curvature tensor $\bar R_{PNKM}$ is bloc-diagonal, i.e. either all  indices are in $\bar{\cal M}_4$  or they are all
in the internal-space directions. For our purposes here, we only need  the curvature tensor with indices in $\bar{\cal M}_4$.
It reads
  \bea
\bar R_{\rho\nu\kappa\mu} =
 k ( \bar g_{\rho\kappa} \bar g_{\nu\mu}  - \bar g_{\rho\mu} \bar g_{\kappa\nu} )
 \  ,
\eea
where $k=-1, 0, 1$ for anti-de Sitter, Minkowski or de Sitter spacetime.
Furthermore, when acting on the factorized perturbation $h_{\mu\nu}(x,y) \equiv h^{\rm [tt]}_{\mu\nu}(x) \psi(y)$,
the wave operator breaks down to a direct sum: $ \bar\nabla^K \bar\nabla_K\,  = \bar \Box^{(2)}_{x} + \bar\Box_y$.
Putting everything together, and using also the four-dimensional wave equation (\ref{massivespin2}),
  leads   to the following linear  variation of the Ricci tensor with indices in  $\{0,1,2,3\}$ :
 \bea\label{Riccifinal}
\hskip -2mm \delta R_{\mu\nu} =  -{1\over 2}  h^{[tt]}_{\mu\nu} \left[ \lambda  - 8k + (2d-4)  (\partial A)^2 +  2 (\bar\Box_y\,  A)
    + (d-2)  (\partial_a  A) \partial^a     +  \bar\Box_y \right]   \psi\, ,
\eea
where internal indices are contracted with the metric $\bar g_{ab}$.
Furthermore, because $h^{[tt]}_{\mu\nu}$ is traceless the variation of the Ricci scalar vanishes,  so that
\bea\label{secondterm}
\delta [g_{\mu\nu} R] =  \, \hat R \, \delta g_{\mu\nu}\,  =
\, h^{[tt]}_{\mu\nu}(x)\,  \psi(y)\,   e^{2A} \hat R\ .
\eea
This completes the computation of the linear variation of the  Einstein tensor in the $\{0,1,2,3\}$ directions.
As the reader can readily verify, the variation of the remaining components of the Einstein tensor vanishes identically.

\vskip 1mm

We turn  next  to the variation of  the energy-momentum tensor  of  ``matter" fields,  i.e. of all
fields other than the $d$-dimensional graviton. These include the scalar and $p$-form gauge fields, which by abuse of notation we will refer
 to collectively as  $\Phi$.   In general, the perturbations $\delta \Phi$  mix at the linearized level
with metric perturbations.  Put differently,  the Lagrangian expanded to quadratic order around the background solution,
may contain terms $\sim \delta g_{MN}\delta \Phi$.
This cannot however happen in the case at hand, because  the special form of metric perturbations and the maximal symmetry of $\bar {\cal M}_4$
  forbid  all contractions of  $\delta g_{MN}$ with any field other than itself.\footnote{This argument does not
  of course exclude cubic or higher-order mixing terms.}
  Indeed,  background fields must be proportional to  invariant AdS$_4$ tensors and neither the metric nor covariant derivatives
  can contract with $\delta g_{\mu\nu}$.
    Thus, it is  consistent to set
 $\delta \Phi=0$ when computing the linearized  equations for spin-2 modes.

\vskip 1mm

 The energy-momentum tensor of the matter fields is given by the variation of their Lagrangian density,
 \bea\label{emt}
T_{MN} \, =\,    {2\over \sqrt{g}} \, {\partial  \over \partial g^{MN}}  (\sqrt{g}\, {\cal L}_{\rm mat}) \ .
\eea
By our previous argument, we need only consider  ${\cal L}_{\rm mat}$  for  $\Phi = \hat \Phi$.
But thanks to the symmetry of the background fields,   ${\cal L}_{\rm mat}(g_{MN}, \hat \Phi) =
 {\cal L}_{\rm mat}({\rm det} (g_{\mu\nu}),  g_{ab}, \hat \Phi)$, i.e. the
Lagrangian density may depend on the 4-dimensional part of the metric only via its determinant.  Indeed,  the only
allowed  background fields with indices in $\{0,1,2,3\}$ are $p$-form field strengths proportional to the $\epsilon$ symbol,
and whose contraction in ${\cal L}_{\rm mat}$ gives a factor of  ${\rm det} (g_{\mu\nu})$.
The reader can also verify that   4D space-time filling sources,  such as D-branes or orientifolds,
do  not modify the above conclusion. It follows
 that at  linear order  $T_{\mu\nu} \propto g_{\mu\nu}$, where  the scalar proportionality factor is made
 out of background fields. Consistency of the trace  relation then implies
\bea\label{linemt}
T_{\mu\nu} = (\hat g_{\mu\nu} + \delta g_{\mu\nu})\, {1\over 4} \, \hat T_\rho^{\ \rho} +  {\cal O} (\delta g^2)\
\   \ ,
\eea
where $ \hat T_{\rho\sigma}$ is the energy-momentum tensor of the background solution.
 By  similar arguments one  shows  that $\delta T_{ab} = \delta T_{a \nu} = 0$ at the  linear level, in agreement with the fact
that the corresponding components of the Einstein tensor  vanish.
\vskip 1mm

To complete our  calculation we now use the
Einstein equations    in order to  express
the partial trace  $\hat T_{\rho}^{\ \rho}$ as follows:
\bea\label{2.2}
 \hat T_\rho^{\ \rho} = \hat R_\rho^{\ \rho} - 2 \hat R
\, =\,
  12 k\, e^{-2A}   - 4e^{-2A} (\bar\Box_y\, A  +
 (d-2)  \, (\partial A)^2)     - 2 \hat R \ ,
\eea
where in the second equality we have used once more the Weyl transformation of the Ricci tensor.
Combining  finally   (\ref{Riccifinal}),  (\ref{secondterm}),  (\ref{linemt}) and   (\ref{2.2}) leads to the following equation
for the wavefunction of spin-2 excitations in the internal space:
\bea\label{eigeneq}
- \left[  \bar\Box_y +  (d-2) \bar g^{ab} (\partial_a  A) \partial_b  \right] \psi  = m^2 \psi \ ,
\eea
where we recall that $m^2 = \lambda - 2k$.
 The above eigenvalue equation  is the main result of this  section.
 It confirms and generalizes the analysis of Csaki et al  \cite{Csaki:2000fc}.
As advertized earlier  the linearized equation only involves the  background metric,   but no other details of the
background fields.
\vskip 1mm

  There are several useful rewritings of this equation.  Firstly,
  in terms of the original internal-space metric  $\hat g_{ab} = e^{2A} \bar g_{ab}$,
 it reads
\bea\label{opmass}
- {e^{-2A}\over \sqrt{[\hat g ]}} \, ( \partial_a  \sqrt{[\hat g]} \,  \hat g^{ab} e^{4A} \partial_b) \,   \psi = m^2 \psi\ ,
\eea
 where  $[\hat g]$ is  the determinant of  $\hat g_{ab}$
 (we reserve the notation $\hat g$ for the determinant of the full $d$-dimensional metric $\hat g_{MN}$).
 Alternatively,  one can put it in the form of a Schr\"ondinger-like problem in the internal space:
 \bea\label{opmassS}
[-\bar \Box_y\,   + V(y) ] \,   \Psi = m^2   \Psi\ ,\qquad {\rm where}\ \  \Psi = e^{{1\over 2} (d-2)A} \, \psi\
\eea
 and the ``analog potential" is given by
\bea\label{Schr}
 V(y) =   e^{-{1\over 2}(d-2)A} \,  \bar \Box_y \, e^{{1\over 2}(d-2)A} \ .
\eea
A third useful rewriting   is as a wave equation in  $d$ dimensions.  One first verifies that  the  identity
 $m^2  = \bar \Box_x^{(2)} -2k =  \bar \Box_x$
is valid  for operators acting on transverse-traceless tensor fields in the
maximally-symmetric spacetime $\bar {\cal M}_4$ (see e.g.  ref. \cite{Polishchuk}).
 The spin-2 equation takes then a  simple and universal  form,  as the scalar  Laplace-Beltrami  (``box") equation in the full $d$-dimensional
 spacetime:
 \bea\label{opmass10}
 {1 \over \sqrt{\hat g }} \, ( \partial_M  \sqrt{\hat g} \,  \hat g^{MN}   \partial_N ) \,   h_{\mu\nu}(x, y) \ =\  0 \ .
\eea

 Note  that the universality of the spin-2 mass operator should be contrasted with what happens for lower-spin fields,
 whose linearized equations typically depend on non-metric  details of the background solution.


\subsection{Normalizability and the zero mode}

The  eigenmode equation  must be supplemented with a  space of admissible
  $\psi(y)$, which requires the introduction of a norm.  To avoid cumbersome notation, let us  first consider the case of a scalar
field and look for factorized solutions of its wave equation of the form
$\delta\Phi(x,y) = \phi(x)\psi(y)$.
Assuming a canonical kinetic term, these solutions are normalized by fixing the following invariant integral:
  \bea
 \vert\hskip -0.3mm\vert \delta \Phi   \vert \hskip -0.3mm \vert^{\, 2}
   \equiv  i \int_{\cal S}  dS^M (\delta\Phi_- \partial_M \delta\Phi_+ - \delta \Phi_+ \partial_M \delta\Phi_-)
\ ,
 \eea
where $dS^M$  is the volume element of  a codimension-one,  spacelike hypersurface ${\cal S}$, and
 $\delta\Phi_\pm$ are the positive- and negative-frequency parts of the solution.
In partcular, for a fixed-time slice\footnote{Strictly-speaking this
 argument applies only to backgrounds with a time-like Killing vector,  either defined globally or at least in an
  asymptotic region where the notion of particles makes sense.}
   in the case at hand, $dS^M =  \sqrt{\hat g}\,  \hat g^{0M} dx^1 dx^2 dx^3 d^{d-4}y$.
   It follows then easily that the norm   factorizes,
  \bea\label{norm}
 \vert\hskip -0.3mm\vert \delta\Phi   \vert \hskip -0.3mm \vert^{\, 2} =  \vert\hskip -0.3mm\vert \phi   \vert \hskip
  -0.3mm \vert_{4}^{\, 2}\times  \vert\hskip -0.3mm\vert \psi   \vert \hskip -0.3mm \vert^{\, 2}\  \qquad
  {\rm where}\qquad
    \vert\hskip -0.3mm\vert \psi   \vert \hskip -0.3mm \vert^{\, 2} \equiv
\int  d^{d-4}y  \sqrt{[\hat g]}\, e^{2A}\,  \vert\psi\vert^2\   \ ,
\eea
and $\vert\hskip -0.3mm\vert  \phi  \vert \hskip -0.3mm \vert_{4}^{\, 2}$ denotes  the norm
of  $\phi(x)$ in the unit-radius  spacetime $\bar {\cal M}_4$.
 The non-trivial feature of the above expression is the warp factor inside the $y$-integral.
 Replacing   $\delta\Phi$   by $\delta g_{\mu\nu}$ introduces two extra inverse  metric factors,
 needed  for the contraction of free indices.
The  extra warp factors  cancel precisely those in $\delta g_{\mu\nu}$, so that  again
  \bea\label{norm1}
\vert\hskip -0.3mm\vert  \delta g  \vert \hskip -0.3mm \vert^{\, 2}  \, =\,
\vert\hskip -0.3mm\vert  h  \vert \hskip -0.3mm \vert_{4}^{\, 2}\, \times  \vert\hskip -0.3mm\vert \psi   \vert \hskip -0.3mm \vert^{\, 2}\
 \eea
where $\vert\hskip -0.3mm\vert  h  \vert \hskip -0.3mm \vert_{4}^{\, 2}$ is   the norm
of  the spin-2 perturbation  $h_{\mu\nu}^{\rm [tt]} (x)$ in  $\bar {\cal M}_4$.
 We conclude that the normalizable excitations are those for which the integral in  (\ref{norm}) is finite.
 When expressed in terms of the wavefunction of the analog Schr\"odinger problem (\ref{opmassS}),
   the norm takes the expected canonical form
  \bea\label{norm2}
 \vert \hskip -0.3mm \vert \psi   \vert \hskip -0.3mm \vert^{\, 2}  =
\int  d^{d-4}y  \sqrt{[\bar g]}\,   \vert\Psi\vert^2\   \ .
\eea

 \vskip 1mm

  One immediate corollary  is that   $m^2$  is non-negative.
 This follows from the form of the  mass-squared operator (\ref{opmass})  and from  the above expression for the norm:
 \bea
 m^2 \, \vert\hskip -0.3mm\vert   \psi   \vert \hskip -0.3mm \vert^{\, 2} =
-\hskip -0.3mm  \int  d^{d-4}y\,   \psi \,  ( \partial_a  \sqrt{[\hat g]} \,  \hat g^{ab} e^{4A} \partial_b) \,   \psi  =
 \int  d^{d-4}y\,   \sqrt{[\hat g]} \,    e^{4A} \vert \partial \psi \vert^2   \geq 0\,  ,
 \eea
 where $ \psi$ is here an eigenfunction with eigenvalue $m^2$.
 The mild assumption in the second step
  is that, upon integrating by parts,   one picks up no
 contributions  either  from singularities or from asymptotic regions.  Conversely,
 since unitarity of the SO(2,3) spin-2 representation
 requires $m^2\geq 0$, a singularity  that leads to a violation  of this  bound must correspond
 to a physically-unacceptable solution.
  \vskip 1mm

 Another simple consequence, modulo the same mild assumption as above,
  is that a massless graviton must necessarily have  $  \psi (y)$ = constant\,
[in which case  the  metric
 perturbation is  proportional to the warp factor]. From the normalizability condition (\ref{norm}) we therefore
 conclude that:
   \bea\label{iff}
  {\rm the \ existence\ of\ a\ massless\ graviton} \ \Longleftrightarrow\  \int  d^{d-4}y\,   \sqrt{[\hat g]} \,    e^{2A} < \infty\ .
  \eea
This condition is, of course,  automatically satisfied for compact internal spaces with smooth warp factor.
It is, however, also a priori compatible with an infinite-volume internal space, as
 in the so-called Randall-Sundrum (RS) model II  \cite{Randall:1999vf}.
``Infinity" in  this model  is, nevertheless, an apparent horizon  \cite{Kaloper:1999sm,Kiritsis:2006ua}
and  the definition of the quantum theory requires a  choice
of boundary conditions.\footnote{This problem  is generic for all flat-brane geometries
with a non-compact transverse space.
Consider indeed an infinite transverse dimension parametrized by a flat coordinate $y$,  and let $e^A$ vanish at infinity.
 For a  particle  moving along  $y$ one has   $ e^{2A} = C \sqrt{e^{2A} - \dot y^2}$\,
  for some constant $C$.  Out at infinity  $\dot y \simeq  e^A \to 0$,    so that
   the  proper time integral  reads:
     \bea
      \int d\tau =  \int dt\, C^{-1}  e^{2A}   \,  \simeq  \,  C^{-1} \int dy\,  e^{A} \ .
    \eea
Geodesic completeness of the  spacetime requires that  this integral diverge,
so that the would-be horizon cannot be reached in finite proper time.
The existence
of a massless graviton requires on the other hand  that  $\int dy\, e^{2A}$ be finite. These two conditions imply that
at infinity $A \simeq  -\nu {\rm log}\, y$ with $1 > \nu > 1/2$.  Such asymptotic behavior is, however, ruled out
by  the weak-energy condition which  requires that  $A^{\prime\prime} \leq 0$  (the ``holographic $c$-theorem)
 \cite{Girardello:1998pd, Freedman:1999gp}. It  can be checked that more than one infinite dimensions cannot help.
}
In the  Karch-Randall model on the other hand,  the  geometry
is geodesically-complete  with   ${\rm AdS}_5$ boundary  regions.  Since the integral
 in (\ref{iff}) diverges at the boundary of
 AdS$_5$, there is no  massless four-dimensional graviton in this model.  This is also the case  for  the smooth interface
 geometries   to which we  turn now our attention.

\section{Interface Solutions and their Charges}
\setcounter{equation}{0}

In this section we review the supersymmetric  solutions of type IIB supergravity
discovered  in references  \cite{DEG1,DEG2}.  We make a choice of coordinates most convenient
for our purposes here, and relate the free parameters of the solutions to
the  charges of the underlying brane configuration
and to the asymptotic values  of the dilaton field.
 For the reader's convenience, we  summarized
in  appendix \ref{Summary of Type IIB supergravity} the  field equations and  Bianchi identities
 of  the type-IIB  bosonic fields,  and  their duality transformations.

\subsection{Local solutions: General form}

  The solutions of interest are fibrations
of  ${\rm AdS}_4\times {\rm S}^2\times  {\rm S}^2$  over a base space which is a Riemann surface $\Sigma$ with the
topology of a disk.  The general discussion of these solutions \cite{DEG2}  is most convenient with a choice of complex coordinate
 that varies over the upper-half plane, but for our purposes here we prefer to use a coordinate that varies over
 the infinite strip:  $$\Sigma \equiv  \{ z \in \mathbb{C}\, \vert \, 0\leq {\rm Im}z \leq {\pi\over 2}\}\ . $$
 The isometry   ${\rm SO}(2,3) \times {\rm SO}(3) \times{\rm SO}(3)$ of the fibers is a
 symmetry of the
 solutions, which also preserve  16 super(conformal) symmetries.
 The solutions considered here  will be  completely specified by two  functions $h_1(z, \bar z)$ and $h_2(z, \bar z)$ which are real
 harmonic and regular inside $\Sigma$, and which obey the boundary conditions (here $\partial_\perp$
 is the normal derivative):
\bea\label{bconditions}
h_1 = \partial_\perp h_2 = 0 \hskip 0.5cm {\rm for} \hskip 0.5cm  {\rm Im}z=0\ ,  \hskip 1cm
h_2 = \partial_\perp h_1= 0 \hskip 0.5cm {\rm for} \hskip 0.5cm  {\rm Im}z= {\pi\over 2}\ .
\eea

In writing down the solutions one also needs the dual harmonic functions,
 which are defined up to a constant by the following  relations
\bea
\label{dualharmfunc}
h_1 = -i({\cal A}_1 - \bar {\cal A}_1) \qquad\rightarrow  \qquad  h_1^D  = {\cal A}_1 + \bar {\cal A}_1\ ,  \no\\
h_2 = {\cal A}_2 + \bar {\cal A}_2 \qquad \rightarrow \qquad  h_2^D = i({\cal A}_2 - \bar {\cal A}_2)\ .
\eea
Note that this ambiguity has a physical interpretation, namely constant shifts of $h_1^D$ and $h_2^D$ correspond, respectively, to gauge transformations of the RR and NSNS two-form gauge potentials.
It is,  furthermore,  convenient to define the following combinations of $h_1$, $h_2$,  and of their first derivatives
(here $\p = \p/\p z, \bar\p = \p/\p\bar z$) :
 \bea\label{W}
W &=& \p  h_1 \bar\p  h_2 + \bar\p h_1 \p  h_2  = \p \bar\p  (h_1h_2)\ ,  \no\\
N_1 &=& 2 h_1 h_2 |\p  h_1|^2 - h_1^2 W \ , \no \\
N_2 &=& 2 h_1 h_2 |\p  h_2|^2 - h_2^2 W   \ .
\eea
In the conventions of appendix \ref{Summary of Type IIB supergravity},
the  supersymmetric solutions of refs.  \cite{DEG1,DEG2} are then described
 by the following backgrounds:
\vskip 0.1mm
\bea\label{metric}
\underline{\rm Metric}:
\qquad ds^2 = f_4^2 ds_{{\rm AdS}_4}^2 + f_1^2 ds_{{\rm S}_1^2}^2 + f_2^2 ds_{{\rm S}_2^{2}}^2 + 4 \rho^2 dz d\bar z \ ,
\eea
where
\bea\label{metricfactors}
f_4^8 &=& 16\, {N_1 N_2 \over W^2}\ ,
\qquad \qquad
\rho^8 = {N_1 N_2 W^2 \over h_1^4 h_2^4}\ ,  \no\\
f_1^8 &=&  16\,  h_1^8 {N_2 W^2 \over N_1^3}\ ,
\qquad  \qquad
f_2^8 = 16\,  h_2^8 {N_1 W^2 \over N_2^3}\ ,
\eea
and the ${\rm AdS}_4$ and
2-sphere metrics are normalized to unit radius;

\bea \label{dilaton}  \hskip -6cm
\underline{\rm Dilaton}:
\qquad  e^{4 \phi} = {N_2 \over N_1}\ ;
\eea
\bea\label{3forms}
\hskip -2.5cm
\underline{\rm 3{\rm -}forms}:  \qquad   F_{(3)}  =   \omega^{\, 45}\wedge db_1  + i \,    \omega^{\, 67}\wedge db_2   \ ,
\eea
where $ \omega^{\, 45}$ and $ \omega^{\, 67}$ are the volume forms of the unit-radius  spheres  ${\rm S}_1^{2}$ and ${\rm S}_2^{2}$,
and \hskip 0.5mm
 \bea\label{3forms1}
b_1 &=& 2 i h_1 {h_1 h_2 (\p  h_1\bar  \p  h_2 -\bar \p  h_1 \p  h_2) \over N_1} + 2  h_2^D \ ,  \no\\
b_2 &=& 2 i h_2 {h_1 h_2 (\p  h_1 \bar\p  h_2 - \bar\p  h_1 \p  h_2) \over N_2} - 2  h_1^D \ ;
\eea
 \bea \label{5form}
\hskip -0.6cm
\underline{\rm 5{\rm -}form}:  \qquad   F_{(5)}  =
 - 4\,  f_4^{4}\,  \omega^{\, 0123} \wedge {\cal F} + 4\, f_1^{2}f_2^{2} \,  \omega^{\, 45}\wedge \omega^{\, 67}
 \wedge (*_2  {\cal F})   \ ,
\eea
where $ \omega^{\, 0123}$ is the volume form of the unit-radius ${\rm AdS}_4$,
${\cal F}$ is a 1-form on $\Sigma$ with the property  that  $f_4^{\, 4} {\cal F}$ is closed,
and $*_2 $ denotes Poincar\' e duality with respect to the $\Sigma$ metric.
 The explicit expression
for ${\cal F}$  is given by
\bea
f_4^{\, 4} {\cal F} = d j_1\   \qquad {\rm with} \qquad j_1 =
3 {\cal C} + 3 \bar  {\cal C}  - 3 {\cal D}+ i \frac{h_1 h_2}{W}\,   (\p  h_1 \bar\p  h_2 -\bar \p h_1 \p h_2) \ ,
\eea
where ${\cal C}$ is defined by the relation $\p {\cal C} = {\cal A}_1 \p  {\cal A}_2 - {\cal A}_2 \p  {\cal A}_1$
while ${\cal D} = \bar {\cal A}_1 {\cal A}_2 + {\cal A}_1 \bar {\cal A}_2$.\footnote{Note that, to match standard
conventions for $\tilde F_{(5)}$, we have introduced an
 additional  factor of $4$ compared to \cite{DEG1,DEG2}. Note also  that the expressions (9.61)
and (9.63) in \cite{DEG1} are missing the factor of ${\cal D}$.}

\vskip 1mm

 The above set of expressions  gives the local form of the general  solution
 for  the ansatz of  refs. \cite{DEG1,DEG2}.
These expressions are invariant under  conformal transformations of the coordinate $z$,
which  map,   however,  in general $\Sigma$ to a different disk-like domain of the complex plane.
The following identities between the dilaton field and the metric factors will prove useful later:
\bea\label{convenientrelations}
f_1^2 f_4^2 = 4 e^{2\phi} h_1^2 \ , \qquad
f_2 ^2 f_4^2 = 4 e^{-2\phi} h_2^2 \ , \qquad
\rho^4 f_1^2 f_2^2 = 4 W^2\ .
\eea
Note  that  the general solutions consistent with the above isometries can also have
a non-vanishing   Ramond-Ramond scalar field.
As has been shown however  in   \cite{DEG1},
it is always possible to set the RR scalar  to zero by a  $SL(2, \mathbb{R})$ duality rotation.
In fact, since
 quantum effects break this
symmetry to  its discrete subgroup $SL(2, \mathbb{Z})$,   continuous  rotations
give  a priori  physically-inequivalent solutions.
Since however only the metric enters in the analysis of the spin-2 spectrum, we wont need
to keep track of  this issue in the present work.


\subsection{${\rm AdS}_5\times {\rm S}^5$ and supersymmetric Janus}

Although any pair of harmonic functions $h_1$, $h_2$  gives a local solution of the supergravity equations,
global consistency imposes stringent restrictions.
 As discussed in  reference \cite{DEG2},  for the solutions of interest
the dilaton and all  metric factors must be regular functions  in the interior  of $\Sigma$, but they may
 diverge  at isolated points of the  boundary.
A further requirement is that  the ${\rm AdS}_4$ scale factor, $f_4$,
 be everywhere non-vanishing, and that (with  the exception of  isolated points) the strip boundary, $\partial\Sigma$,
should correspond  to  interior points of the ten-dimensional geometry.
\vskip 1mm

 Removing the singular points of $f_4$ separates $\partial\Sigma$ into a collection of boundary segments.
 The last regularity condition then  implies that
on each segment  one of the  two sphere radii  must vanish.
 This is  guaranteed by   (\ref{bconditions}), as   can be  checked  with the help of
 (\ref{convenientrelations}).
To avoid a conical singularity, we need however also to  fix the rate at which the
 sphere radii vanish.
 Explicitly, for ${\rm Im} z = \epsilon$ and  ${\rm Im} z = \pi/2 -  \epsilon^\prime$
 with $\epsilon,  \epsilon^\prime \ll 1$, we must require that
   \bea
 {f_1^2\over 4\rho^2}\ \simeq \ \epsilon^2 \qquad {\rm and}  \qquad
  {f_2^2\over 4\rho^2} \ \simeq \  \epsilon^{\prime\, 2}  \ .
 \eea
  Further conditions may still be needed in order to constrain the
 allowed singularities at isolated points on the boundary of $\Sigma$.   Admissible singularities
 include  those that  can be interpreted as coming from  known string-theory  branes.

\vskip 2mm
A simple choice for  the  harmonic functions,     consistent with all
the  above physical  requirements,  is as follows:
\bea\label{hAdS}
h_1 =  -i \alpha_1\, {\rm sinh} z  +  {\rm c.c.}    \qquad {\rm and} \qquad
h_2 =  \, \alpha_2 \,  {\rm cosh}z  + {\rm c.c.}   \   ,
\eea
where $\alpha_1$ and $\alpha_2$ are arbitrary real parameters.  After a little algebra\footnote{The following
identities are useful: $(c+\bar c)(s-\bar s) = (c\bar s - \bar c s)(1+c\bar c + s\bar s)$ ,
$(c+\bar c)(c\bar s - \bar c s) = (s-\bar s)(1+c\bar c - s\bar s)$
and $(s-\bar s)(c\bar s - \bar c s) = (c+\bar c)(1-c\bar c + s\bar s)$,  where $c\equiv {\rm cosh}z$, $s\equiv{\rm sinh}z$
and bars are complex conjugates.}
 these give:
\bea
\rho^4 =   \vert  \alpha_1\alpha_2\vert \ ,  \qquad \qquad \ &&  \qquad
f_4^2 =   4\rho^2  \cosh^2 (\frac{z + \bar z}{2})  \ ,
\no\\
f_1^2 =  4 \rho^2 \,   \sin^2(\frac{z - \bar z}{2i})  \ ,
 \qquad \ &&  \qquad
f_2^2 =  4 \rho^2 \,    \cos^2(\frac{z - \bar z}{2i}) \ .
 \eea
 Writing
$z = x+ i y$, one recognizes  immediately the  ${\rm AdS}_5 \times {\rm S}^5$ metric,
\bea
ds^2 = L^2 \left[ dx^2 + \cosh^2(x) ds^2_{{\rm AdS}_4} + dy^2 + \sin^2(y) ds^2_{{\rm S}^2_1} +  \cos^2(y)   ds^2_{{\rm S}^2_2} \right]\
\eea
with radius   $L^2  = 4 \vert  \alpha_1\alpha_2\vert^{1/2}  $.  The choice (\ref{hAdS}) leads furthermore to
 vanishing 3-form fluxes, and to the  following constant dilaton
and 5-form backgrounds:
\bea
e^{2\phi}  = \left\vert {\alpha_2\over \alpha_1} \right\vert \qquad {\rm and}\qquad
F_{(5)} = -4  L^4 (1 + \ast) \omega^{\, 4567y}\ ,
\eea
where $\omega^{\, 4567y}$ is the volume form on the unit ${\rm S}^5$,  with orientation fixed by the
sign of $\alpha_1\alpha_2$.  The radius of the 5-sphere is related as usual  to
the total  $D3$-brane charge:
\bea
\int_{S^5} F_{(5)} = -4 \mbox{Vol}(S^5) = -4 \pi^3 L^4\ .
\eea
All the solutions that we will discuss will be  obtained by deforming the harmonic functions
of this basic ${\rm AdS}_5\times {\rm S}^5$ solution.

\vskip 2mm
A one-parameter family of deformations, the supersymmetric Janus configurations \cite{DEG1}, is obtained
by the following trick: one notes that a translation along the strip preserves  the boundary conditions
of the harmonic functions.   A common translation of the arguments of $h_1$ and $h_2$  is
 of course   just  a  reparametrization.
New regular solutions can, however,  be found by a relative translation, i.e.
\bea\label{hJanus}
h_1 =  -i \alpha_1\, {\rm sinh} (z - {\Delta\phi\over 2} )  +  {\rm c.c.}    \qquad {\rm and} \qquad
h_2 =  \, \alpha_2 \,  {\rm cosh}(z+ {\Delta\phi\over 2})   + {\rm c.c.}   \   ,
\eea
where $\Delta\phi$ is a real parameter (whose name will be justified in a moment).
 The regularity of the Janus solution has been established in ref. \cite{DEG1}.
 In the asymptotic regions  $x\to \pm \infty$  we expect it to approach the basic
 ${\rm AdS}_5 \times {\rm S}^5$  geometry. To find the asymptotic values of the radius and  dilaton
 requires some care, because $W$ vanishes at the leading exponential order.  A simple calculation
 actually gives
   \bea\label{WJanus}
 W =   \, - \alpha_1\alpha_2   \, {\rm cosh} (\Delta\phi) \, {\rm sin} (2y)\ ,
 \eea
 so the deformation simply rescales  $W$ by a factor ${\rm cosh} (\Delta\phi)$.
 Using this fact,   and keeping the leading exponential behavior for all other terms  in  the expressions
 (\ref{W}) to (\ref{dilaton}),
 leads to  the following values for the dilaton  and  the radius
 of    the two asymptotic  ${\rm AdS}_5 \times {\rm S}^5$ regions,  at $x\to\pm\infty$:
 \bea\label{asRadius}
L^4 =  16 \vert\alpha_1\alpha_2\vert {\rm cosh} (\Delta\phi)  \qquad {\rm and}\qquad
e^{2\phi_\pm} = \left\vert {\alpha_2\over \alpha_1} \right\vert e^{\pm \Delta\phi}\ .
\eea
 The  Janus configuration   describes  a  smooth  supersymmetric
 domain wall,  across which the   dilaton  changes by a total  amount $\Delta\phi$.
 The wall lives in the near-horizon geometry of D3-branes. This simplest
 solution has no five-brane charge,  as is obvious from the fact that there is no non-contractible 3-cycle
 that can  support it.

\subsection{Non-vanishing five-brane charge}
\label{5chargesection}

Adding five-brane charge to the supersymmetric Janus solution requires  harmonic functions
with  singularities
at the  boundaries of the strip. These singularities describe new asymptotic regions
of the ten-dimensional geometry.\footnote{As explained in reference \cite{DEG2} each five-brane
singularity can be  split into a pair of singularities of branch-cut type like those describing the
asymptotic AdS$_5\times$S$^5$ regions.  Note that after such a splitting, there may appear conical singularities in the bulk of $\Sigma$.
To see this, note that in the upper-half-plane coordinates the function $W$ and correspondingly the $\Sigma$-metric, $\rho$, vanish at isolated points in the bulk of $\Sigma$.  It has not been explicitly checked whether these points lead to conical singularities or not.}
Proceeding in steps,  we consider first the following two-parameter deformation of Janus:
\bea\label{hNS5}
h_1 &=&  -i \alpha_1\, {\rm sinh} (z - \beta_1)  +  {\rm c.c.} \ ,  \no\\
h_2 &=&  \, \left[ \alpha_2 \,  {\rm cosh}(z-\beta_2)   -    \gamma\,  {\rm ln}({\rm tanh} {z\over 2})\right]  +  {\rm c.c.}   \  .
\eea
All the parameters in these expressions must be real.
The function ${\rm ln}({\rm tanh} {z\over 2})$ is purely imaginary for Im$z = {\pi\over 2}$,   so $h_2$
continues to vanish on the upper boundary of the strip. It also  obeys  a Neumann  condition on the lower
boundary, except at the origin, $z=0$,  where it   has a
logarithmic   singularity. This  is not very surprising,   since we expect
 the metric and the dilaton to be singular
 at the position of  a  5-brane.  \vskip 1mm
\vskip 1mm

To show why the above solution has  NS5-brane charge, we need to
identify a non-contractible
3-cycle that can support
 3-form flux.  Consider an open  curve,  ${\cal I}$,  starting and ending on the lower  boundary of the infinite strip.
Since  ${\rm S}_1^2$  shrinks to zero size at both ends,   ${\cal I}$ and this 2-sphere fiber
gives a 3-cycle with the topology
of a 3-sphere.  When $\gamma=0$ this cycle is contractible,  because  ${\cal I}$  can  shrink  continuously
  to a point.  For $\gamma\not =0$ on the other hand, a curve with its endpoints on either side of $z=0$
  cannot be contracted.  There is no RR 3-form flux through this
  cycle  because, as can be seen from eq.\,(\ref{3forms}),
  the RR 3-form  is proportional to the volume of  the second sphere,
  ${\rm S}_2^2$.\footnote{The reader can
  check that  the integral of ${\rm Im}F_{(3)} $ over ${S^2_2 \times {\cal I}}$ does not vanish, so the solution (\ref{hNS5})
  has a non-zero RR 3-form field. But  $S^2_2 \times {\cal I}$ is a cycle with boundaries and,  in accordance with the absence of
  D5-brane charge,  the integral of ${\rm Im}F_{(3)}$  over any closed 3-cycle is zero.}
 The NS-NS flux through ${\rm S}_1^2\times {\cal I}$,  on the other hand,  reads
 \bea\label{h2dual}
Q_{\rm NS5} = \int_{{\cal I } \times S^2_1} {\rm Re}F_{(3)} = \int_{\cal I} d b_1 \int_{S^2_1} \omega^{45} =   4 \pi b_1 |_{\p \cal I} = 8\pi h_2^D  |_{\p \cal I}\ ,
\eea
where $\p \cal I$ is the boundary of  ${\cal I}$.
The last of the above equalities follows from eq.\,(\ref{3forms1}),  and the  vanishing of  $h_1$ on  the  real-$z$  axis.
 The function dual to $h_2$ is
 \bea
 h_2^D  =  \, \left[ i \alpha_2 \,  {\rm cosh}(z-\beta_2)   -   i \gamma {\rm ln}({\rm tanh} {z\over 2})\right]  +  {\rm c.c.}   \  ,
\eea
from which it follows easily that $  h_2^D  |_{\p \cal C} = 2\pi\gamma$.
Note that the entire contribution to the charge comes from the imaginary
part of the logarithm, and only depends on the topological class of the open curve ${\cal I}$, i.e. on whether its two endpoints lie
on the same or on opposite sides of $z=0$. Inserting in (\ref{h2dual}) gives the
 relation between $\gamma$ and the NS5-brane  charge,
\bea\label{5charge}
Q_{\rm NS5} = 16\pi^2\, \gamma\ .
\eea

\vskip 1mm

 To relate the remaining four parameters of (\ref{hNS5}) to physical parameters of the solution,
 we consider again the asymptotic geometry in the two limits  $x\to\pm\infty$.
 Note first  that the extra term in the expression for $h_2$ has the asymptotic behavior
 \bea
- \gamma {\rm ln}({\rm tanh} {z\over 2}) \, \simeq\,   \Bigl\{
\begin{array}{cc}   2\gamma e^{-z} \hskip 5mm  &{\rm when} \ x\to\infty\ ,   \\
-i\gamma\pi + 2\gamma e^{z}  \hskip 3mm  &{\rm when} \ x\to-\infty\ .  \\
\end{array}
\eea
Adding this to the hyperbolic cosine gives
\bea
h_2 \simeq  \alpha_2^{\pm} \ {\rm cosh} (z - \beta_2^{\pm}) \,  +\,   {\rm c.c.}   \hskip 8mm  {\rm when}\  \ x\to\pm \infty\ ,
\eea
where
 \bea\label{phys1}
 \alpha_2^\pm = \alpha_2\, \sqrt{1 + {4\gamma\over \alpha_2}e^{\mp \beta_2}}\ ,
 \qquad  e^{\beta_2^\pm}  = e^{\beta_2}\, (1 + {4\gamma\over \alpha_2}e^{\mp \beta_2})^{\pm1/2} \ .
 \eea
By our previous argument, we may  now compute the radius  and the value of the dilaton
in  the two asymptotic ${\rm AdS}_5 \times {\rm S}^5$ regions with the result:
 \bea\label{phys2}
L_\pm^4 =  { Q^\pm_{{\rm D}3}\over 2\pi^3}\,  =\,
 16 \vert\alpha_1\alpha_2^\pm \vert \, {\rm cosh} (\beta_1 - \beta_2^\pm)
 \qquad {\rm and}\qquad
e^{2\phi_\pm} = \left\vert {\alpha_2^\pm \over \alpha_1} \right\vert e^{\pm (\beta_1 - \beta_2^\pm)}\ .
\eea
Notice that the asymptotic radii and D3-branes charges need not be the same at the two ends, as was the case in
the Janus solution. Indeed,  the D3-branes can either {\it  intersect } or {\it end}
 on the  NS5-brane stack,  which is allowed to carry D3-brane charge  \cite{Strominger:1995ac}.
 The fraction of D3-brane charge dissolved inside the NS5-branes is
 \bea
 {Q^-_{{\rm D}3}- Q^+_{{\rm D}3}\over Q^+_{{\rm D}3}  + Q^-_{{\rm D}3} } \ = \
 {2\gamma \sinh \beta_1 \over 2\gamma \cosh\beta_1 + \alpha_2 \cosh (\beta_1-\beta_2)}\ .
 \eea
This vanishes when $\gamma=0$, since in this case there are no NS5-branes,  but also when $\beta_1=0$
which corresponds to all D3-branes intersecting the NS5-branes. The other extreme,
where this ratio approaches one, corresponds either to $\gamma\to\infty$, or to
 $\beta_1, \beta_2\to \infty$ with the  difference $\beta_1 -\beta_2$ held fixed.
In these limits  the NS5-brane stack  absorbs the  totality of D3-brane charge in its worldvolume.



  \vskip 1mm

 A solution with  D5-brane charge can be obtained from eq. \hskip -1mm (\ref{hNS5}) by
 an  S-duality
 transformation. The  harmonic functions transform as a doublet
 of SL(2, $\mathbb{Z}$), so that under this transformation
$h_1\rightarrow h_2$ and  $h_2\rightarrow -h_1$. The transformation also exchanges the two 2-spheres.
 To restore the initial boundary conditions (\ref{bconditions}), we
 furthermore make the (analytic)  reparametrization of the strip  $z\to i\pi/2 - z$.
  The D5-brane background  corresponds thus, finally,  to the functions
 \bea\label{hD5}
h_1 &=&  \, \left[ i\alpha_2 \,  {\rm sinh}(z+\beta_2)   +    \gamma {\rm ln}\left({\rm tanh} ({i\pi\over 4} - {z\over 2})
\right)\right]  +  {\rm c.c.}  \ ,  \no\\
h_2 &=&   \alpha_1\, {\rm cosh} (z + \beta_1)  +  {\rm c.c.}   \  .
\eea
It can be easily checked that the metric of this solution is the same as that of the NS5 brane, while
 the sign of the dilaton is  flipped, and the NSNS and RR 3-forms
are  exchanged. This is precisely the action of the S-duality element of
   SL(2,$\mathbb{Z}$).

 \vskip 2mm

It is now easy to  write down a  solution, which involves both NS5-brane
  and D5-brane stacks   wrapping  the two different 2-spheres of the background geometry.
  The corresponding harmonic functions read:
 \bea\label{hNS5D5}
h_1 &=&    \, \left[ -i \alpha_1\, {\rm sinh} (z - \beta_1) -    \gamma_1 {\rm ln}\left(  {\rm tanh} ({i\pi\over 4} - {z - \delta_1\over 2})\right)
\right]
 +  {\rm c.c.} \ ,  \no\\
h_2 &=&  \, \left[ \alpha_2 \,  {\rm cosh}(z-\beta_2)   -    \gamma_2 {\rm ln}\left({\rm tanh} ({z - \delta_2\over 2})\right)\right]  +  {\rm c.c.}   \  .
\eea
This solution describes the near-horizon geometry of stacks of  intersecting D3-branes,  NS5-branes  and D5-branes,
whose worldvolumes in  the asymptotically-flat  spacetime are along the
  directions (0123), (012456) and (012789). The five-brane charges are
 \bea
 Q_{\rm D5} = 16\pi^2\gamma_1\ , \qquad
 Q_{\rm NS5} = 16\pi^2\gamma_2\ .
 \eea
 The asymptotic values of  the ${\rm AdS}_5 \times {\rm S}^5$ radii and the dilaton field are :
 \bea
 L_\pm^4\,  =\,
 16 \vert\alpha^\pm_1\alpha_2^\pm \vert \, {\rm cosh} (\beta^\pm_1 - \beta_2^\pm)
 \qquad {\rm and}\qquad
e^{2\phi_\pm} = \left\vert {\alpha_2^\pm \over \alpha^\pm_1} \right\vert e^{\pm (\beta^\pm_1 - \beta_2^\pm)}\ ,
 \eea
 where $\alpha_j^\pm$ and $\beta_j^\pm$ for $j=1,2$ are given by:
  \bea\label{phys2}
 \alpha_j^\pm = \alpha_j\, \sqrt{1 + {4\gamma_j\over \alpha_j}e^{\pm(\delta_j -  \beta_j) }}\ ,
 \qquad  e^{\beta_j^\pm}  = e^{\beta_j}\, (1 + {4\gamma_j\over \alpha_j}e^{\pm(\delta_j - \beta_j) })^{\pm1/2} \  .
 \eea
 The ratio  $L_-^4/L_+^4$ determines,  as previously,   the fraction of  D3-branes dissolved into the two stacks of 5-branes.
 The fractions dissolved separately  into the D5-brane stack,  and the NS5 stack,
 can be  computed from the integrals of the RR  5-form over ${\cal I}\times S^2_1
 \times S^2_2$,  where $\cal I$ is a semicircle around the corresponding logarithmic singularity on the strip boundary.

\vskip 1mm
 There is, actually, a well-known subtlety in the definition of D3-brane charge  due to the Chern-Simons term
 of the type-IIB action, as reviewed nicely in reference  \cite{Marolf:2000cb}.  The gauge-invariant and  conserved  {\emph {Maxwell charge}}
   differs  in general from the {\emph {Page charge}}  which (being quantized and localized)  is the one that counts
 the number of D3-branes. In the asymptotic AdS$_5$ regions the two notions of charge coincide, so that  their difference, i.e. the  total number
 of D3-branes dissolved in the five-brane stacks,  is  unambiguous.  The separate fractions of D3-branes going into  the NS5-stack and the
 D5-stack may suffer from a  gauge ambiguity, but we will not  need  their  explicit
 expressions here.
 We just note that the total parameter count is correct:  there are
 seven physical parameters,  the two asymptotic values of the dilaton,  the two 5-brane charges, and the three D3-brane charges,
i.e. the numbers of D3-branes that are   (i)  dissolved  in the
 NS5-branes,  (ii)  dissolved in  the D5-branes,  or  (iii)  intersecting both  stacks.
   This is the same as  the number of parameters in (\ref{hNS5D5}),   if one excludes   the spurious parameter
  that  fixes the origin of the real-$z$ axis.

\vskip 1mm
  The solution (\ref{hNS5D5}) can be generalized further by including any number of five-brane singularities,
  both in the upper and in the lower boundaries of the strip. These describe  configurations with different fractions of D3-branes
  dissolved into more than two distinct  five-brane stacks. They can be analyzed by a straightforward extension
  of the above discussion.


\section{Localized Gravity: what to look for ?}
\setcounter{equation}{0}

In this section we briefly review the  Karch-Randall model  \cite{KR1},   and  in particular
those  features of its prototype thin-brane geometry  that are responsible for the  localization of  gravity.
 A key qualitative feature of this geometry is that the warp factor has a local bump, whose height  generates a
 large scale hierarchy. As we will see, the Janus configuration does not share this qualitative feature,   and
 it cannot therefore localize gravity.  The reason is that a dilaton domain wall has no a priori prescribed
  tension and thickness, because its superpotential vanishes.
Solutions with both NS5- and D5-brane
 charge are, in this regard, more promising, but their detailed analysis
will be  postponed to a forthcoming companion paper \cite{tocome}.

\subsection{The Karch-Randall model}

  The starting point of the Karch-Randall model  is the  effective 5D  thin-brane action
   \bea\label{IKR}
   I_{\rm KR} \, = \,    - {1\over 2 \kappa_5^{\, 2}} \int d^4x \, dy \,  \sqrt{g} \left(  R +  {12 \over L^2} \right)  +  \lambda \int d^4x \, \sqrt{ [g]_4}\ ,
  \eea
  where $[g]_4$ is the determinant of the induced   metric at $y=0$.
 This action  depends on three dimensionful parameters: the 5D
 gravitational coupling $\kappa_5$, the tension $\lambda$ of the thin 3-brane, and the
  radius   of curvature  $L$ of the bulk  spacetime.  The important  dimensionless combination  is \
  $ \kappa_5^{\ 2} \lambda L$,  or equivalently the length parameter
  \bea
  y_0  = L \, {\rm arctanh} \left( {\kappa_5^{\ 2} \lambda L \over 6}\right) \ ,
  \eea
  in terms of which the solution of the Einstein equations can be written as  \cite{Kaloper:1999sm,Cvetic:1993xe,DeWolfe:1999cp}
  \bea\label{2AdS5}
  ds^2 = L^2 {\rm cosh}^2 \left( {y_0 - \vert y\vert \over L}\right) \, \bar g_{\mu\nu} dx^\mu dx^\nu + dy^2\
  \eea
with   $\bar g_{\mu\nu}$  the metric of  the unit-radius AdS$_4$.
The  spacetime (\ref{2AdS5}) consists  of two identical slices of AdS$_5$ of radius $L$,
glued by  Israel matching conditions at the location of the thin 3-brane.
  The parameter  $y_0$
is the turn-around point of the warp factor, and it moves to infinity as one tunes \hskip 0.3mm    $ \kappa_5^2 \lambda L \to 6$
(see figure \ref{fig:warp-KR}).

\begin{figure}[htb]
\centering
\includegraphics[width=0.5\textwidth]{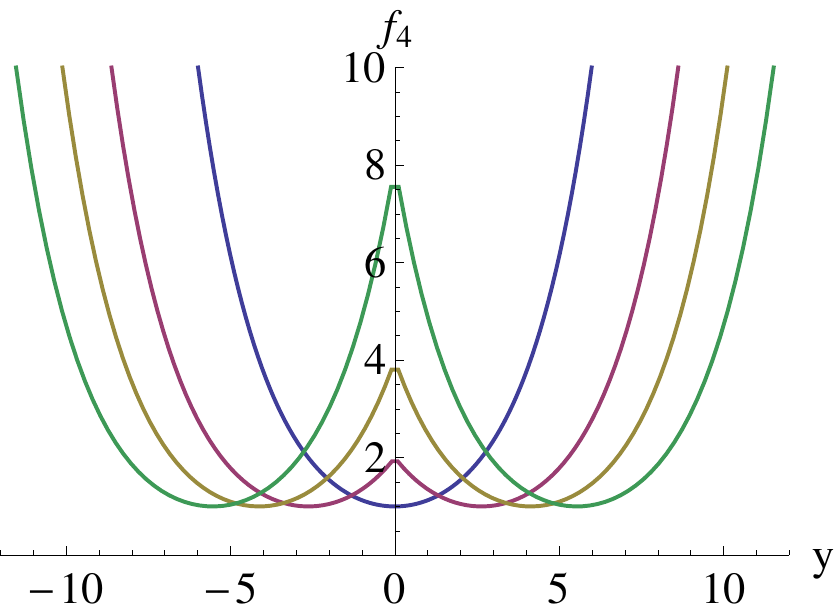}
\caption{\footnotesize The AdS$_4$ warp factor ($f_4$) of the Karch-Randall geometry with $L=1$
for $l = 1, 2, 4$ and $8$ (blue, purple, yellow, and green, respectively).}
\label{fig:warp-KR}
\end{figure}

\vskip 1mm
  The physical  parameters in four dimensions  are   the radius  of curvature $\ell$
  of the AdS$_4$ brane-world,   Newton's constant $G_N$, and   the characteristic Kaluza-Klein length,
  $\ell_{\rm KK}$,  below which Newton's
  force law is modified by  5D physics.
  Overlooking the fact that our actual universe is  de Sitter, not anti-de Sitter, we  have
  the following observational requirements (for short-range tests of gravity see e.g. ref. \cite{Adelberger:2003zx}) :
 \bea\label{obs}
 \sqrt{ \ell\,  \ell_{\rm Planck}}  \sim 1 \ {\rm  mm} \ , \qquad
 {\ell \over \ell_{\rm Planck}}  \sim 10^{62} \ , \qquad {\rm and}\  \  \ell_{\rm KK}  <  1\, {\rm mm}\ .
 \eea
 To convert these to constraints on the 5D parameters, one first notes that
  \bea
  \ell^2 =  \, e^{2A(0)}\, =\,  L^2 \, {\rm cosh}^2 \left( {y_0 \over L}\right)\ ,
\eea
 so for the brane-world to be almost flat we need $y_0$ bigger
 (but not hierarchically bigger)  than $L$. The geometry near the 3-brane can then be approximated by
 the exponential warp factor of the Randall-Sundrum model \cite{Randall:1999vf},
   \bea\label{2AdS5c}
  ds^2\  \simeq\   \ell^2\,  e^{ - 2{\vert y\vert / L}} \, \bar g_{\mu\nu} dx^\mu dx^\nu + dy^2\  .
  \eea
The role of the geometry beyond the turn-around points $y_0$ is essentially to replace the adhoc
boundary condition at the  AdS$_5$ horizon  by a suitable normalizability condition.
The main conclusions of reference \cite{Randall:1999vf} can then be carried over to this bent-brane case.
  The effective 4D Newton's constant, in particular, reads
  \bea\label{GN}
8\pi G_N \, = \, \kappa_5^2 \, e^{2A(0)}\, \vert \psi_0(0)  \vert^2\
  \simeq \,  \kappa_5^2 \, e^{2A(0)} \left(  \int_{-y_0}^{y_0}  dy \, e^{2A} \right)^{-1} \,  \simeq \,  \kappa_5^2 / L   \ ,
\eea
where $\psi_0 (y)$ is the wavefunction of the  lightest, nearly-massless graviton mode.
The corrections to Newton's law  from the exchange of higher excitations are given by
  \bea\label{deltaV}
   V_{\rm Newton}   + \Delta V  \  \simeq \  - {G_N m_1 m_2\over r}\, (1 +   \gamma  {L^2\over  r^2} + \cdots ) \   ,
 \eea
with $\gamma$ a numerical constant.  Note that the relation between  $G_N$ and the 5D coupling is
the same as for  a standard Kaluza-Klein circle of  size $L$, but the corrections to Newton's law in the two cases are
different.  Either way,   in view of the experimental constraint (\ref{obs})
 we need  $L$ to be less than (a fraction of)  the millimeter.
 \vskip 1mm

  A key feature of the above prototype geometry is the existence of a local maximum of the warp factor
 near $y\simeq 0$.  Its immediate vicinity supports the nearly-constant,  nearly-massless  mode $\psi_0$, whose mass
 has been shown  in ref. \cite{Miemiec:2000eq} to be given by  $ m_0^2  \simeq {3} L^2/2 \ell^2$.  This is of the same order
 as the blueshift between the local maximum at $y=0$ and the minima at $y=\pm y_0$.
 The estimate for the  wavefunction in (\ref{GN})
 is based on the assumption that $\psi_0(y) $
is almost constant  at $y \ll y_0$,   so  that the contribution to its norm vanishes exponentially fast away  from
the brane.

The large blueshift between the points $y=0$ and $y=y_0$  is crucial for suppressing the contribution of the excited
modes. These have  AdS$_4$ masses  $m \sim 1$, so that they mediate interactions  whose range is comparable to   the
radius of curvature, $\ell$,  of the braneworld.  Their support is, however, centered around $y_0$, and they have
very small  amplitude at the position $y=0$
of the 3-brane.  The actual calculation (see e.g.  \cite{Randall:1999vf,Brandhuber:1999hb, Garriga:1999yh,
Kehagias:1999my, Kiritsis:2002ca,Callin:2004py}
and references therein) gives $\vert \psi (0 \vert m) \vert^2 \sim m L^2/\ell^2$, so that
the contribution to the gravitational force,
 \bea\label{GNc}
{\delta V } \  = \   V_{\rm Newton}\times \, \vert \psi_0(0)\vert^{-2}  \,  \sum_{\{m\not= m_0 \}}  \vert \psi (0 \vert m) \vert^2 \, e^{-m r/\ell} \ ,
\eea
takes, at distances  $L \ll r \ll \ell$,   the subleading form exhibited in equation (\ref{deltaV}).
The precise form of the  KK correction (which also includes contributions from non-TT modes) will,
in general,  depend on details of the geometry. What appears, however, to be essential for the localization of gravity is
 the very large suppression of the excited modes at the location of the brane.

\vskip 1mm

Some further intuition on this issue can be  derived from the equivalent Sch\"odinger problem
defined in eqs. (\ref{opmassS})  and  (\ref{Schr}). Using a coordinate in which the rescaled internal
metric is constant  ($\bar g_{55} = e^{-2A} \hat g_{55} = \ell^{-2}$) one finds
 \bea
-{d^2 \Psi \over dz^2} +\left[  {15 \over 4(L^2 + z^2)}   - {3\over L} \delta (z) \right] \, \Psi\,
 \simeq\,  {m^2\over  \ell^2}  \Psi
\qquad {\rm where} \ \ 1+ {\vert z\vert \over L}   \simeq  e^{  {\vert y\vert / L}}   \ .
\eea
These expressions are actually  valid for $z\ll \ell$, while the exact potential turns around and
diverges at some large but finite value $z_0$   \cite{KR1}. The key feature of this analog
potential is its volcano shape, which traps the nearly massless graviton near $z\sim 0$, and expells
from this region the excited modes. In previous studies of
  thick-brane models   \cite{DeWolfe:1999cp, Csaki:2000fc, Bazeia:2007nd}
this shape was  engineered by  appropriately tuning an adhoc superpotential of  scalar fields.
One important issue that   concerns us here is whether such a shape can be obtained
in string theory.

 \vskip 1mm

 We complete this section with a comment on  gravity  localization as seen from the  perspective
 of the dual defect field theories.   As explained in ref. \cite{Aharony:2003qf}, the mode expansion of supergravity fields
 is mapped to the expansion of conformal  operators in the 4D bulk, as they approach
  the three-dimensional defect. A nearly-massless localized
 AdS$_4$ graviton should thus correspond to an operator ${\cal O}_0$ with  dimension $\Delta \simeq  3$   in the
 expansion of the energy-momentum tensor.  Furthermore, the  behavior of the corrections to Newton's
 force in the  Randall-Sundrum  model corresponds to the  $r^{-4}$  fall-off of the
   two-point function of the bulk energy-momentum tensor \cite{Duff:2004wh}.\footnote{ To touch basis with the usual discussion
in Poincar\' e coordinates, note that the
distance $r$ between two points  on the  AdS$_4$ brane
  is related to the geodesic  distance $R$  in the embedding AdS$_5$ spacetime  through
  $L^2 ( {\rm cosh}^2 {R \over L}-1) = \ell^2 ( {\rm cosh}^2 {r \over \ell}-1)$.
 Removing the standard flat-brane cutoff in the AdS/CFT correspondence amounts to zooming into the region
  $L\ll r \ll \ell$ . } It is also interesting to note that, while the 3-point functions of the energy-momentum tensor in a homogeneous
  CFT are completely fixed by Ward identities, this is not the case for the 3-point functions involving  the reduced operator ${\cal O}_0$.
  This reflects the fact that our hypothetical nearly-massless graviton need not have universal couplings to all the forms
  of matter.\footnote{We thank J. Penedones and S. Rychkov for a discussion of this point.}


\subsection{Warp factors and a limiting geometry}
\label{limitinggeometry}

   It has been argued in reference  \cite{KR2} that intersecting D3-branes  and D5-branes  could provide a string-theory embedding of
  the Karch-Randall model. The underlying
  hypothesis was that the back-reaction of the D5-branes on the AdS$_5\times$S$^5$ geometry could be
  well approximated by the effective thin-brane action (\ref{IKR}). Given the explicit solutions of D'Hoker et al \cite{DEG1, DEG2}
  one can try to find out   whether this assumption  was  justified. The questions  to ask are:
  (1) does the warp factor  develop a ``local bump"  with a large  hierarchy of scales?  and (2) can the
   KK modes, both on the strip and on the two-spheres,    be kept  within the experimental limits?
   The second question is of course harder to address, since it  depends on the detailed shape of the analog potential
   as well as the metric $\bar g_{ab}$  in the internal space.

  \vskip 1mm

   The warp factor in  the solutions of D'Hoker et al reads
\bea
e^{2A} = f_4^2 =  2 \, ({N_1N_2\over W^2})^{1\over 4}\ ,
\eea
where $W, N_1, N_2$ have been defined in eq. (\ref{W}). We first consider the case of Janus, for which the harmonic
functions (\ref{hJanus}) can be written as
  \bea
 &h_1 =  2 \alpha_1\cosh (x- {\Delta\phi \over 2})\, \sin y  \ , \qquad h_2 =  2 \alpha_2 \cosh (x + {\Delta\phi \over 2})\, \cos y  \ ,
 \eea
 where  $z =x+iy$.
 A simple calculation then  gives
 \bea\label{Wjanus}
& h_1 h_2 =   \alpha_1\alpha_2  \sin 2y\,  ({\rm cosh}\Delta\phi +  \cosh2x) \ ,  \qquad \hskip -2mm
 W  =  -  \alpha_1\alpha_2  \sin 2y\,   {{\rm cosh}\Delta\phi } \  ,
\eea
and after a slightly more tedious algebra:
\bea\label{squareb}
 & N_1 =  \alpha_1^3 \alpha_2\, \sin 2y\,  \Bigl[   \sinh \Delta\phi\, \cos 2y\,  \sinh (2x +  \Delta\phi)
  -    \sinh \Delta\phi \, \sinh 2x \, ( \cosh 2x + 2 \cosh \Delta\phi )      \nonumber \\
& + \cosh \Delta\phi\,  \left( 1 +   \cosh^2 2x + 2 \cosh 2x \cosh \Delta\phi  \right)
\Bigr]      \ ,
\nonumber \\
\ &   \nonumber
 \\
 & N_2 =  \alpha_1 \alpha_2^3\, \sin 2y\,  \Bigl[   \sinh \Delta\phi\, \cos 2y\,  \sinh (2x -  \Delta\phi)
  +    \sinh \Delta\phi \, \sinh 2x \, ( \cosh 2x + 2 \cosh \Delta\phi )      \nonumber \\
& + \cosh \Delta\phi\,  \left( 1 +   \cosh^2 2x + 2 \cosh 2x \cosh \Delta\phi  \right)
\Bigr]      \ .
\eea
  As a check  note that for $\Delta\phi = 0$
the   square brackets in the above expressions  reduce to $[4 \cosh^4 x]$, so that the warp  factor
is $y$-independent. This is indeed  the AdS$_5\times$S$^5$  limit in which the geometry factorizes.
   \vskip 1mm

For  $\Delta\phi \not= 0$  the square brackets in (\ref{squareb}) depend  on the coordinate $y$,
and the geometry does not factorize.  Note in passing  the symmetry under the reflection  $(x, y) \to (-x, \pi/2 - y)$,
 which exchanges the harmonic functions $h_1$ and $h_2$
and thus leaves the  metric invariant. Along  the  middle line of the strip, i.e. at  $y=\pi/4$,  the wrap factor is given
up to an overall constant  by
$$
e^{8A} \propto \left( 1 +   \cosh^2 2x + 2 \cosh 2x \cosh \Delta\phi  \right)^2 - {\rm tanh}^2\Delta\phi \sinh^2 2x ( \cosh 2x + 2 \cosh \Delta\phi )^2 \ .
$$
Figure  \ref{fig:warp-Janus}  shows $e^A \equiv f_4$  as a  function of the invariant distance
 $X =   \int_0^{\pm x}  2\rho dx$, for different values of the parameter $\Delta\phi$.
  As  these plots show, and as one can   check  explicitly from the above expression,
 $f_4$  is an even function of $x$  increasing monotonically from the  center out  to the AdS$_5$ boundary regions.
The absence of a local bump  reminiscent of the  thin-brane geometry of \cite{KR1}  is a clear indication that Janus
 cannot possibly localize gravity.

\begin{figure}[]
\vspace{-1cm}
\begin{center}
\begin{tabular}{c @{\hspace{0.5in}} c}
\includegraphics[height=2.0in]{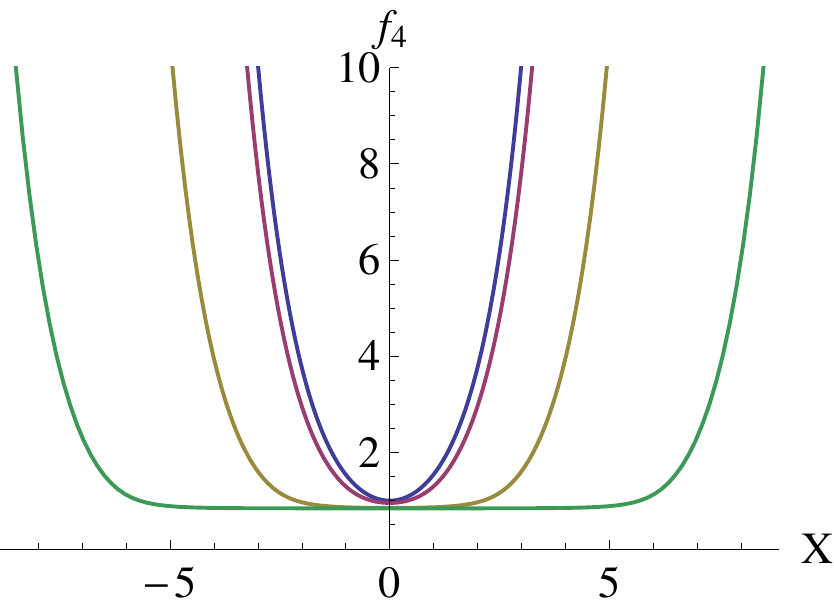} &
\includegraphics[height=2.0in]{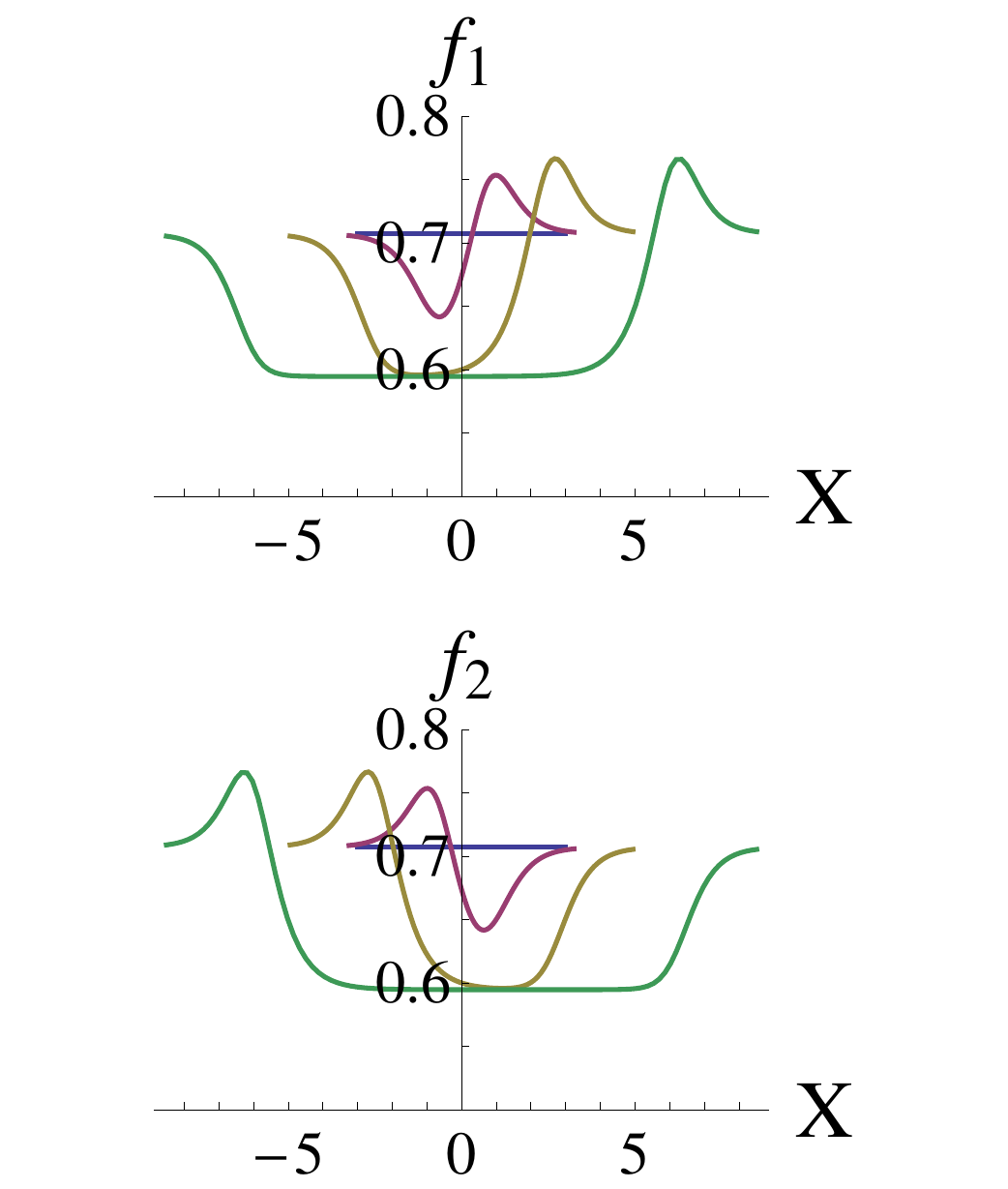}
\end{tabular}
\end{center}
\caption{\footnotesize   The AdS$_4$ warp factor ($f_4$)  and the sphere radii ($f_1, f_2$) of the Janus geometry
for dilaton jumps $\Delta\phi = 0, 1, 4$ and $10$ (blue, purple, brown and  green curves, respectively).
The horizontal axis gives the invariant distance from the strip center,
along the line $y= \pi/4$.  The asymptotic AdS$_5$ radii have been set to one, and the  dilaton in the left
asymptotic region has been set  to zero. Note the absence of a local maximum of the warp factor.}
\label{fig:warp-Janus}
\end{figure}

 One other thing seen in figure \ref{fig:warp-Janus} is that the AdS$_5$ trap becomes broader and flatter
 as the dilaton jump $\Delta\phi$ is tuned  up. Interestingly, the limit $\Delta\phi\to \infty$ leads to
 a smooth, geodesically-complete geometry. Indeed, for $\vert x\vert \ll {\Delta \phi}$  one gets
 \bea
 &N_1 \simeq {1\over 4} \alpha_1^3  \alpha_2\, e^{2\Delta\phi} e^{-2x} \sin 2y (1+ 2\sin^2y)\ ,  \nonumber \\
\ &  \nonumber  \\
&N_2 \simeq {1\over 4} \alpha_1 \alpha_2^3\, e^{2\Delta\phi} e^{2x} \sin 2y (1+ 2\cos^2y)\ ,
\eea
leading to the following expressions  for the metric and the dilaton:
 \bea
 ds^2 =   L^2 (3+ \sin^2 2y)^{1/4} \left[ {1\over 2} ds^2_{{\rm AdS}_4} + dz d\bar z + ({\sin^2y\over 1+2\sin^2y}) ds^2_{{\rm S}^2_1}
 + ({\cos^2y\over 1+2\cos^2y}) ds^2_{{\rm S}^2_2}
 \right]\ ,   \nonumber
 \eea
 \vskip -2mm
 \bea
  e^{2\phi} = \Bigl\vert {\alpha_2\over\alpha_1} \Bigr\vert \left( {1+2\cos^2y\over 1+2\sin^2 y }\right)^{1/2} e^{2x} \  .
 \eea
 This limiting   geometry describes therefore  a five-dimensional,  non-compact
 spacetime  (AdS$_4\times \mathbb{R}$)
  with a  dilaton varying linearly in  $ \mathbb{R}$,   and a
 warp  factor  varying   smoothly over a deformed 5-sphere. For large but  finite $\Delta\phi$
 the nearly-flat fifth dimension has size $\sim  \Delta\phi$, and the
  Kaluza-Klein excitations should therefore have  characteristic
spacings  $\delta m \simeq ({\Delta \phi})^{-1}$. This will be indeed
 confirmed by the numerical analysis of the Janus spectrum  in section \ref{sec:numspec}.

 \vskip 1mm

 The reason for this behavior of the Janus solution can be understood if one notes that
 the  dilaton has no superpotential at the classical level. This should be contrasted with
 toy models of thick brane worlds  \cite{Csaki:2000fc,DeWolfe:1999cp,  Bazeia:2007nd},
 where the superpotential can be engineered to produce the desired warp factor.
The tension and thickness of scalar-field domain walls  are, in particular,   tunable parameters in these
 toy models. Dilaton domain walls, on the other hand, would tend to spread to infinite thickness and zero tension
 in the absence of gravity.
 They are only stabilized in the case at hand by  the asymptotic AdS$_5$ regions.

 \vskip 1mm

\begin{figure}[t]
\vspace{-1.5 cm}
\centering
\begin{tabular}{c @{\hspace{0.5in}} c}
\includegraphics[width=0.4 \textwidth]{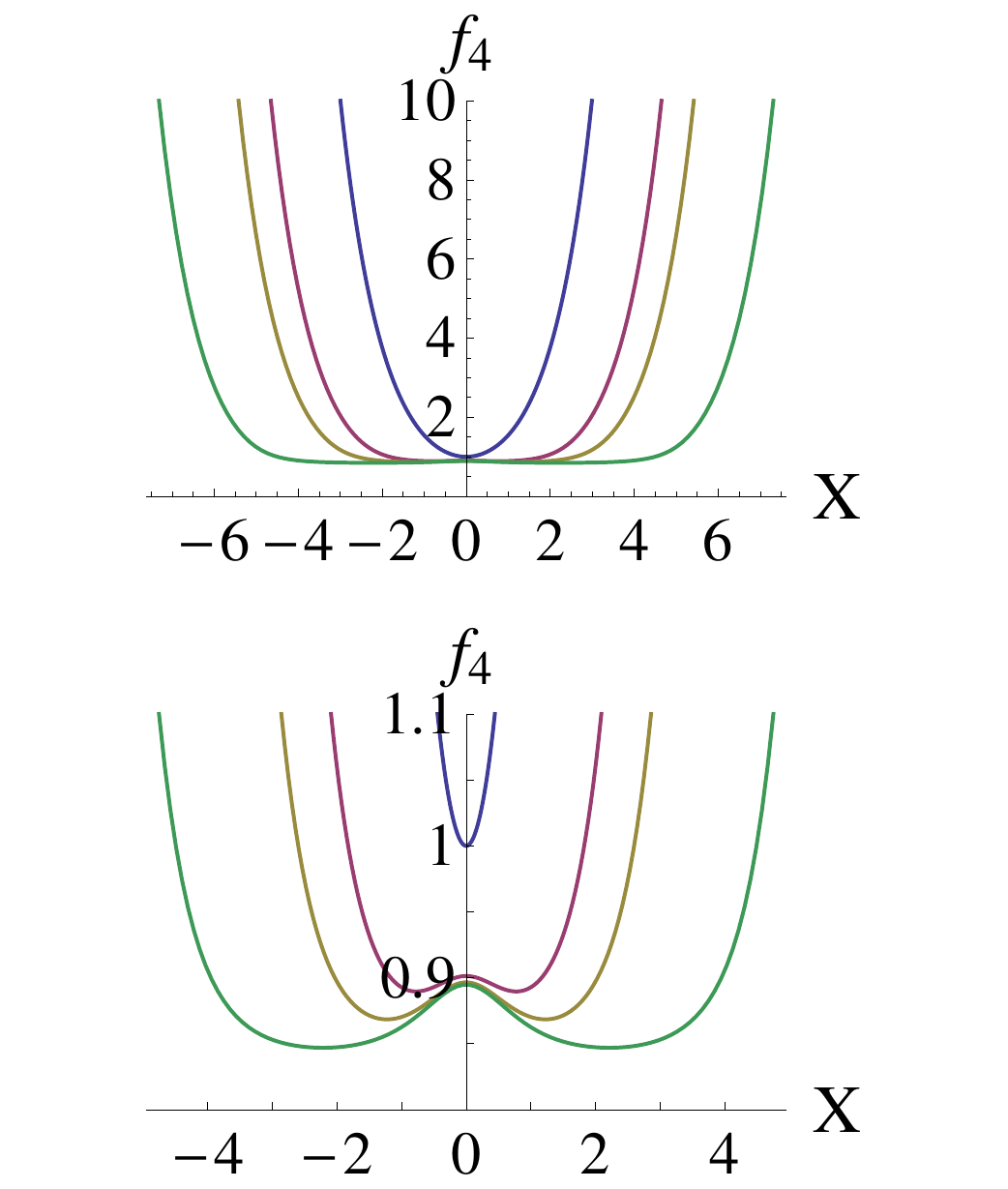} &
\includegraphics[width=0.4 \textwidth]{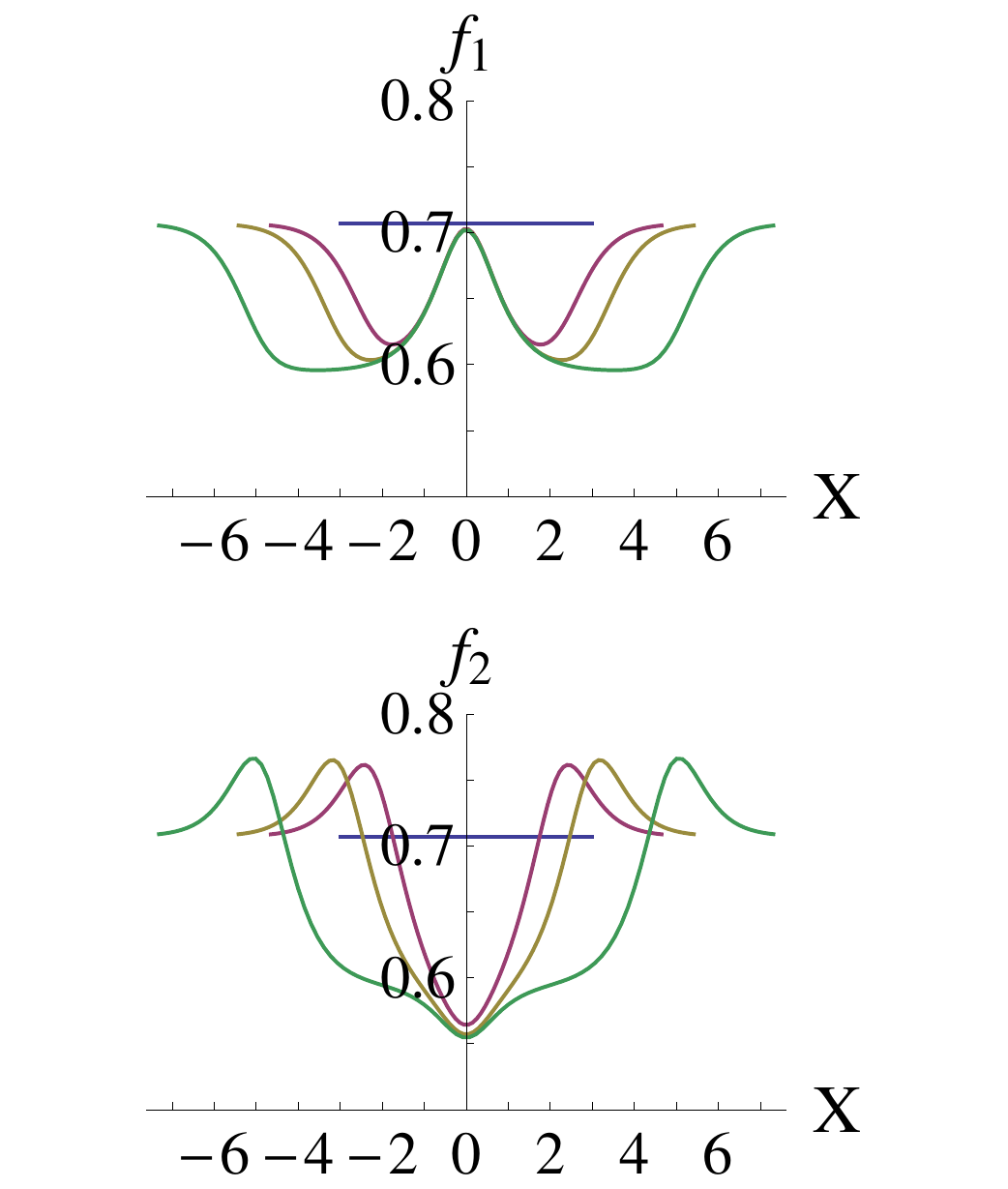}
\end{tabular}
\caption{\footnotesize
The warp factor along the $y = \pi/4$ line for an NS5-brane stack (see section \ref{5chargesection}), with  charge $Q_{NS5} = 16 \pi^2 \gamma$
 where $\gamma = 0,1/2,1,5$ (blue, purple, yellow and  green curves, respectively).  The horizontal axis gives the invariant distance from the strip center.  The asymptotic AdS$_5$ radius has been set to one, and the asymptotic dilaton has been set to zero.  Note, as can be seen in the lower left figure, that the local maximum of $f_4$ does
 not exceed  the asymptotic value of the AdS$_5$ radius.
}
\label{fig:warp-5brane}
\vspace{5mm}
\end{figure}
\begin{figure}[h!]
\centering
\begin{tabular}{c @{\hspace{0.5in}} c}
\includegraphics[width=0.4 \textwidth]{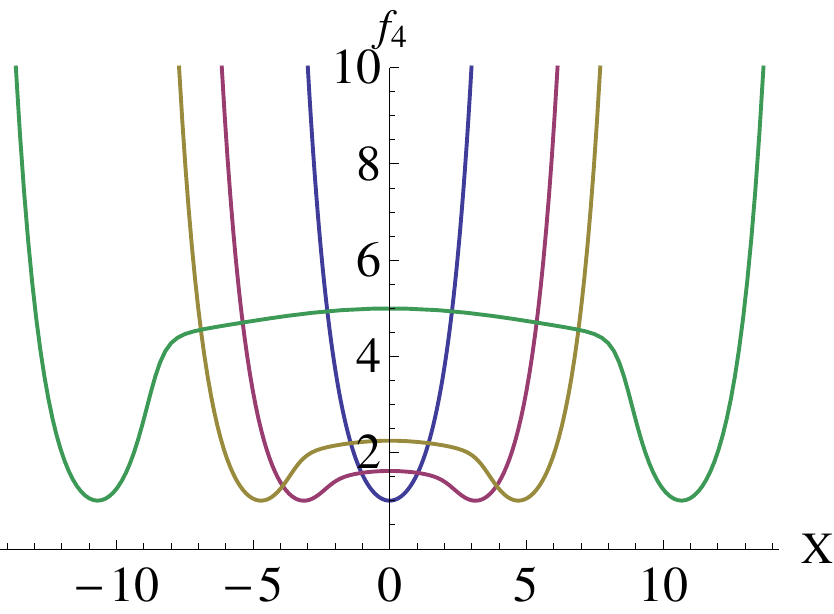} &
\includegraphics[width=0.4 \textwidth]{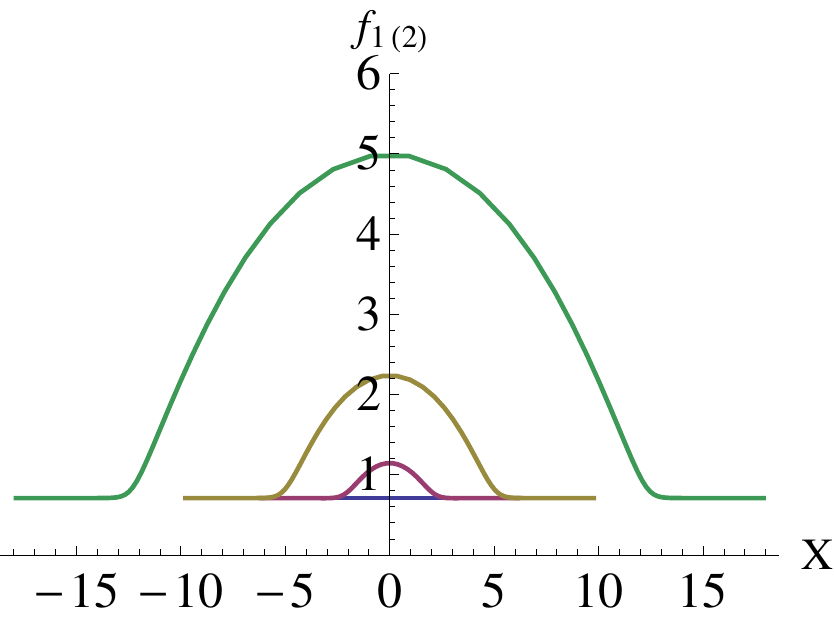}
\end{tabular}
\caption{\footnotesize
Warp factor along the $y = \pi/4$ line  for intersecting D5-brane and NS5-brane stacks (see section \ref{5chargesection}),
with brane charges $Q_{D5} = Q_{NS5} = 16 \pi^2 \gamma$ where $\gamma = 0,1/2,1,5$ (blue, purple, yellow and  green curves, respectively).
 The conventions are the same as in figure \ref{fig:warp-5brane}.
   Note that on the line $y = \pi /4$
   the $S^2$ warp factors are identical ($f_1 = f_2$).}
\label{fig:warp-55brane}
\end{figure}

\vskip 5mm

  Five-branes are more promising in this respect,   since they are stable
   in  flat  spacetime.  Figure \ref{fig:warp-5brane}  depicts the AdS$_4$ warp factor and the sphere radii for  solutions
   with NS5-brane charge, while  solutions with  both NS5-brane and D5-brane charge
   are shown in figure \ref{fig:warp-55brane}.  As can be seen from these figures, the warp factor does  indeed exhibit   a
   characteristic  bump,  but   this grows  wider and flatter as it gets taller.  It thus looks as if a
    scale hierarchy
   cannot be arranged without decompactifying the internal space.  This question will be
   analyzed in detail in ref. \cite{tocome}.
   For now,  we conclude   with a simple observation: while
    the warp-factor bump is tiny in figure \ref{fig:warp-5brane},  it is  much more pronounced in figure \ref{fig:warp-55brane}.
     The reason is that the NS5-brane stack (whose tension in the Einstein frame scales  as $e^{\phi/2}$)
      pushes the dilaton field to a larger value,  thereby reducing its back-reaction on the local geometry.
     The D5-branes (whose tension scales like $e^{-\phi/2}$)
     exerts an opposite pressure on the dilaton.
      When there are both NS5-branes and D5-branes the dilaton reaches an equilibrium value,
      and the total tension cannot be further  reduced.

 \section{Spectral Problem on the Strip}
 \label{sec5}
\setcounter{equation}{0}


We go back now to the  eigenmode equation  of section 2, and
specialize it to the interface solutions of D'Hoker et al. We will see, in particular,
that  for excitations that are constant on the 2-spheres,  the
 operator  $(m^2 + 2)$ is  the  Laplace-Beltrami operator on the strip $\Sigma$,  with
metric  $ (2 \rho/ f_4)^{2} \, dz d\bar z$ and with Neumann boundary conditions.
As a check, we will diagonalize this operator to recover the well-known AdS$_5\times$S$^5$
mass spectrum. In the next section we will  analyze  the first non-trivial example, that of
the supersymmetric Janus solution.


\subsection{Reduction of the eigenmode equation}

The eigenvalue equation (\ref{opmass}) for spin-2 modes depends only, as we have seen,  on the
warp factor and the internal-space metric.  For the interface solutions of  D'Hoker et al
these can be read from the general expression (\ref{metric}),
 \bea
A  = {\rm ln} f_4\ , \qquad  \hat  g_{ab}\,  dy^a dy^b\,   = \,
f_1^2 ds_{{\rm S}_1^2}^2 + f_2^2 ds_{{\rm S}_2^{2}}^2 + 4 \rho^2 dz d\bar z    \ .
\eea
Exploiting the background isometries, we may, furthermore, expand the wavefunction  $\psi$  in a basis of spherical harmonics, i.e.
in self-explanatory  notation
\bea
\psi (y^a) =    Y_{l_1 m_1} Y_{l_2 m_2} \, \psi_{l_1l_2}(z,\bar z)\ ,
\eea
where we have suppressed the dependence of $Y_{lm}$ on the sphere coordinates.
Inserting these expressions in (\ref{opmass}) leads to the following eigenvalue equation:
 \bea\label{interfacespectrum}
 - f_4^{\, 2} \left[ {\delta^{ij}\over 4\rho^2} \partial_i\partial_j
+  {\delta^{ij}\over 2\rho^2} ( \partial_{i} \ln (f_4^2 f_1  f_2 )) \partial_j
-  {l_1(l_1+1) \over  f_1^{\ 2}} -{l_2(l_2+1)\over  f_2^{\ 2}}
\right] \hskip -0.2mm  \psi_{l_1l_2} = m^2  \psi_{l_1l_2}
\eea
with $z\equiv  y^1+iy^2$   the complex coordinate of the strip,   and $i,j=1,2$.
The lowest mass eigenvalue corresponds to a wavefunction with $l_1=l_2=0$.
 Much  of our analysis will be restricted to this class of wavefunctions.

\vskip 1mm

The  $l_1=l_2=0$ equation takes a simple form if expressed in terms of the harmonic functions $h_1$
and $h_2$. This follows from the identities (\ref{convenientrelations}),
  which imply in particular
  \bea\label{ids2}
f_4^2 f_1 f_2 =  4 \vert h_1 h_2\vert \   ,   \qquad  {\rm and}\qquad
 \frac{\rho^2}{f_4^2}  =   \left\vert \frac{W}{2 h_1 h_2}\right\vert  \,  =\  - { \p\bar\p (h_1h_2) \over 2 h_1h_2}    \ .
 \eea
 In the second equation we  have used the fact  that $W/h_1h_2$ is  a
 negative-definite real function  in the interior of $\Sigma$. From the above identities one gets
 \bea
0 &=&     \p  \bar\p  \psi_{00} \,
+\left( \p  \ln (f_4^2 f_1 f_2)\right)  \bar \p \psi_{00} + \left( \bar\p  \ln (f_4^2 f_1 f_2)\right)  \p \psi_{00} \, +\, m^2\,  { \rho^2\over f_4^2} \, \psi_{00}
\no\\ && \no\\
&=&  \p  \bar\p \psi_{00} \,
+ \left(\p \ln (h_1 h_2)\right) \bar \p\psi_{00}   + \left(\bar\p \ln (h_1 h_2)\right)  \p\psi_{00} \, - \,  m^2\,  { W  \over  2 h_1h_2}    \psi_{00}
 \ .
 \eea
Multiplying now by $(h_1h_2)$  and   rearranging various  terms puts the eigenvalue equation in the following elegant form:
 \bea\label{besteq}
  { 2 h_1h_2 \over  W}
  \,  \p\bar\p\,   \tilde\psi_{00} \ =\    (m^2 +2)  \tilde\psi_{00} \ ,  \qquad {\rm where}
 \qquad  \tilde\psi_{00} \equiv  h_1h_2\psi_{00}\ .
\eea
The operator $(m^2+2)$ for this class of $s$-wave excitations  is thus a   Laplace-Beltrami operator on the
 strip, with the effective
metric  $ (2 \rho/ f_4)^{2} \, dz d\bar z$. Note, as a check,  that a constant  wavefunction $\psi_{00}$
implies $\tilde \psi_{00} = h_1h_2$,  which solves the above equation
 for zero mass  in accordance with the general  expectations.
 This would-be massless graviton is not,  however,
  a normalizable mode, as will become clear in a minute.

\vskip 1mm
 The  reduced equation (\ref{besteq}) is manifestly  invariant under the $S$-duality transformation, which exchanges
$h_{1}$ with $h_{2}$.  This should be so,  because $S$-duality  does not affect  the (Einstein-frame)  metric.
Independent rescalings of $h_1$ and $h_2$,
   which change the values of the radius and the dilaton  in the asymptotic  ${\rm AdS}_5\times {\rm S}^5$ regions,
   are also symmetries of the above eigenvalue problem.
This is  a consequence of the scale covariance of the two-derivative supergravity action, and of the special form of the universal
 dilaton coupling.
\vskip 1mm

For completeness, let us also write down  the equation for arbitrary values of the angular momenta, $l_1$ and $l_2$,
 on the two spheres:
  \bea\label{besteq1}
 \left[  {  2 h_1h_2 \over  W}
  \,  \p\bar\p\,  - \, {N_1\over W h_1^3 h_2}\, l_1(l_1+1)\, - \, {N_2\over W h_2^3 h_1}\, l_2(l_2+1)\,
    \right] \,  \tilde\psi_{l_1l_2} \ =\    (m^2+2 )  \tilde\psi_{l_1l_2} \ ,
 \eea
 where $ \tilde\psi_{l_1l_2} = h_1h_2 \psi_{l_1l_2}$ as above.This is again $S$-duality invariant,
 if one exchanges  $l_1$ with $l_2$.
Note  that  $Wh_1h_2$ is negative-definite within the strip,  so as expected  non-vanishing angular momenta
add a positive term to the mass-squared operator.

\vskip 1mm
    To fully specify the spectral problem, we must also discuss the space of allowed functions  on the strip.
Recall that generic points on the strip boundary are regular interior points of the 10-dimensional geometry,
and ${\rm Im} z = \epsilon\to 0$ is the radial polar coordinate of a local 3-dimensional coordinate patch.
Requesting that the metric perturbation be analytic in the local  Cartesian coordinates  therefore
  implies
\bea\label{bc1}
 \psi_{00} = {\rm constant} + {O}(\epsilon^2)\ , \qquad {\rm and} \ \  \psi_{l_1l_2} = {O}(\epsilon)\ \  {\rm if}\ \ l_1+ l_2\not= 0 \ .
\eea
Thus $\psi_{00}$ has a Neumann boundary condition on the real axis, while for all other $\psi_{l_1l_2}$  the condition
 is  Dirichlet.
The same conditions apply to the upper boundary of the strip. As for the behavior near singular points and at infinity, this
is fixed by requiring normalizability of the wavefunction.
From the general expression (\ref{norm}) for the norm  one finds
\bea
\vert\hskip -0.3mm\vert  \psi   \vert \hskip -0.3mm \vert^{\, 2}  \, =  \,
    \int_{\Sigma}  d^2z\,   (4\rho^2 f_1^2 f_2^2 f_4^2)\,  \vert \psi_{l_1l_2}\vert^2\ ,
\eea
where the integration runs over the strip $\Sigma$.
Using  the identities (\ref{ids2})  one arrives then at the following reduced expression:
\bea\label{bc2}
\vert\hskip -0.3mm\vert  \psi   \vert \hskip -0.3mm \vert^{\, 2} \
=\ \int_{\Sigma}  d^2z\,  {\vert W h_1h_2\vert }\, \vert \psi_{l_1l_2} \vert^2\
= \int_{\Sigma}  d^2z\,  {\bigl\vert {W\over  h_1h_2} \bigl\vert }\, \vert \tilde\psi_{l_1l_2} \vert^2\
  .
\eea
This norm guarantees that the Laplace-Beltrami operator is self-adjoint.
 The integral must, in particular, converge in the asymptotic AdS$_5\times$S$^5$ regions, as well as in the vicinity
 of   five-branes.   The boundary conditions   (\ref{bc1}) together with the above norm  determine the space of allowed
  spin-2 excitations in the background of all  interface solutions.

\vskip 1mm

  We mention for future reference the equivalent Schr\"odinger problem (\ref{opmassS} - \ref{Schr}), which
  is defined on the full six-dimensional internal space with metric
  \bea
  \bar g_{ab}\,  dy^a dy^b  &=&   f_4^{-2} \left(  f_1^2 \, ds_{{\rm S}_1^2}^2 + f_2^2\,  ds_{{\rm S}_2^{2}}^2 + 4 \rho^2\,  dz d\bar z \right)
  \no\\ && \no\\
  &=& \left\vert {h_1^2 W\over N_1}\right\vert \, ds_{{\rm S}_1^2}^2 + \left\vert {h_2^2 W\over N_2}\right\vert \, ds_{{\rm S}_2^2}^2
  + \left\vert {2 W\over h_1 h_2}\right\vert \, dz d\bar z\ .
  \eea
  Though the analog potential $V(y) = f_4^{-4}  \, \bar \Box_y  f_4^4$  can help  in some cases one's  intuition,
  we  will not use it in the present work.   The reduced form of the equation will be more convenient for our purposes here.


\subsection{The case of  AdS$_5\times$S$^5$}

To verify our formulae, and as a simple warm-up
exercise,  let us  first analyze  the pure  ${\rm AdS}_5\times {\rm S}^5$ geometry. Writing
$z\equiv x+iy$,   one has:
\bea
\frac{W}{2 h_1 h_2} =  - \frac{1}{4 \cosh^2x}\  ,  \qquad  {\rm and}\ \ \
h_1 h_2 =  2 \alpha_1\alpha_2 \cosh^2x \sin 2y\  \ .
\eea
We focus on the wavefunctions $\psi_{00}$ which, as we have just seen,  must obey Neumann boundary conditions
on the infinite strip.  Their $y$-dependence can thus be decomposed in the basis of functions   $\{$cos$(2ny) \}$  for  $ n= 0,1, 2 \cdots $.
From the form  (\ref{besteq}) of the wave operator, one guesses the following form of orthogonal eigenfunctions:
\bea\label{separation}
\psi_{00}(x,y) =   \,  e^{2in y}
\,  (\sum_{k= 0}^n e^{-4i k  y}) \,  \chi_n(x)\  \Longrightarrow\  \tilde\psi_{00}(x,y)\, = \,  \sin \left[ 2(n+1)y\right]\
  \tilde \chi_n(x)\,   ,
 \eea
where $\tilde \chi_n(x) \equiv  2 \alpha_1 \alpha_2 \,  \cosh^2x\, \chi_n(x)$.  With this ansatz
  the eigenvalue equation reduces to an ordinary differential equation
for  the function $\tilde \chi_n(x)$,
\bea\label{Leg1}
  -  \cosh^2\hskip -0.3mm x\,  \left[ {d^2\over d x^2} \,         -    4(n+1)^2 \right] \, \tilde \chi_n \ = \   (2 + m^2) \,  \tilde \chi_n     \ .
\eea
This is the Legendre equation, as can be recognized after the change of variables ${\rm tanh} x \equiv  t$ which
brings it to the  canonical form  \cite{mag}
 \bea
\left[ {d\over dt} (1-t^2) {d\over dt} - {4(n+1)^2 \over 1-t^2} +   \kappa (\kappa + 1)  \right] \, \tilde \chi_n = 0 \, , \ \  {\rm where} \ \
  \kappa (\kappa + 1) \equiv 2+m^2  .
\eea

\vskip 1mm
The general solution of the Legendre equation is a linear combination
\bea
\tilde \chi_{n,\kappa}(t)  \, =\,  c_1 P_{\kappa}^{2n+2} (t) + c_2 Q_{\kappa}^{2n+2} (t) \ ,
\eea
where $c_1, c_2$ are arbitrary coefficients,  and
$P_{\kappa}^{2n+2}$ and $Q_{\kappa}^{2n+2}$ are the associated Legendre functions of, respectively,  the first
and  the second kind.  To find the normalizable eigenmodes, note that in
terms of the   variable $-1< t <1$  the  criterion (\ref{bc2}) reads
  \bea
  \int_{-1}^1 dt\   \vert \tilde \chi_{n,\kappa} \vert^2 \ <\ \infty\ .
  \eea
 At $t\to 1$ the Legendre function  $Q_{\kappa}^{2n+2}$ has a pole  of order $n+1$, while $P_{\kappa}^{2n+2}$ has
 a zero of the same order \cite{mag}.  Convergence of the integral in this limit requires therefore that
   $c_2=0$. At   $t\to -1$, on the other hand,  the function
 $P_{\kappa}^{2n+2}$ has a pole of order $n+1$, but with residue proportional to $\sin(\kappa \pi)$.
 To avoid the pole  we must thus choose  $\kappa\in \mathbb{Z}$. Since $\kappa$ and $-\kappa - 1$ correspond to the same
 value of the mass, we may actually restrict  $\kappa$ to the natural integers. Finally,
Rodrigues' formula
\bea
 \tilde  \chi_{n,\kappa}\  \propto\  P_{\kappa}^{2n+2} (t) \, =   \,
 {1\over 2^{\kappa} \kappa !}\, (1-t^2)^{n+1}  {d^{\kappa + 2n+2}\over dt^{\kappa + 2n+2}}\,  (t^2-1)^{\kappa}\ ,
\eea
shows that when $\kappa < 2n+2$ our would-be solution is identically zero.  This leads then to the
following  spectrum of normalizable modes:
\bea\label{specAdS}
m^2 = (\kappa - 1) (\kappa + 2)\,  , \qquad {\rm for}\ \   \kappa \ {\rm integer} \geq 2n+2 \, ,  \  {\rm and}\ n=0,1,2 \cdots\  \ .
\eea
The lowest-lying spin-2 excitation is  as expected massive,  with  $m^2 = 4 $.
\vskip 2mm

    The above spectrum corresponds to  eigenmodes which are constant on both 2-spheres of the background metric (\ref{metric}).
In the case at hand these 2-spheres make up,   together with the $y$-interval,   the 5-sphere of the AdS$_5\times$S$^5$ solution.
Thus all modes can be  organized  in representations of the bigger symmetry group  $SO(6)$.
 As can be easily checked,  the states (${n,\kappa}$)  belong  to
the symmetric tensor product of $n$  vector representations of $SO(6)$.  Their partners in this  same
 representation, for  $n\not=0$,  have a non-trivial
dependence on the coordinates of the 2-spheres, which is given by the 5-sphere harmonics.

The $n=0$ states are $SO(6)$ singlets,  and there is one such state for each
 $\kappa = 2,3 \cdots$. This decomposition of the AdS$_5$ graviton into AdS$_4$ eigenstates
agrees with  the five-dimensional analysis of Karch and Randall \cite{KR1}.
States with the same $n$ and different $\kappa$ are related by  conformal, $SO(2,4)$ transformations.
Note, however,  also that for  given $\kappa$ there are $[\kappa/2]$ different  $SO(6)$
representations with equal mass. This degeneracy is not accidental -- it is due to  the fact that
 all  graviton states belong to a single unitary irreducible
representation  of the supergroup $PSU(2,2\vert 4)$.



\section{Spectrum of Supersymmetric Janus}
\setcounter{equation}{0}

We  will now study the   spectrum of  spin-2 excitations in the supersymmetric
Janus background.  The key observation is that the $l_1=l_2=0$ eigenmode equation is in this case separable, and it reduces to
  an ordinary differential equation   with   four regular singular points.
  We will solve this equation numerically,
 and exhibit the spectrum as a function of the dilaton-jump parameter $\Delta\phi$.
 The result confirms the expectations from section 4, in particular in the limit of  large
 $\Delta\phi$  in which the interface spreads out to
    a  large nearly-flat extra dimension with a linearly-varying  dilaton.



\subsection{Heun's equation}

Although the Janus geometry does not factorize, it
 follows from eqs. (\ref{Wjanus}) that  the ratio  $(W/h_1h_2)$   is $y$-independent.
 Thus, at least for the $l_1=l_2=0$ modes,  we can use the ansatz (\ref{separation})
 to separate the variables $x$ and $y$. The eigenvalue equation
(\ref{besteq}) reduces then to the following ordinary differential equation:
  \bea\label{Heun1}
  -  \,  \frac{{\rm cosh}\Delta\phi +   \cosh2x}{2\, {\rm cosh}\Delta\phi }
  \,   \left[ {d^2\over d x^2} \,         -    4(n+1)^2 \right] \, \tilde\chi_n \ = \   (2 + m^2) \, \tilde \chi_n     \ ,
\eea
\hskip 1mm
\noindent   \hskip -1.1mm  where here $\tilde\chi_n \equiv \alpha_1\alpha_2
({{\rm cosh}\Delta\phi +   \cosh2x})\, \chi_n$.
  Changing again to the variable $t\equiv\tanh x$,  and defining
the convenient ``Janus parameter"
 \bea
   ({\rm coth}{\Delta\phi\over 2})^2\ \equiv \ \xi   \in (1, \infty ]  \ ,
 \eea
brings the above equation to  the   equivalent  form  \vskip -2mm
  \bea\label{Heun2}
\left[ {d\over dt} (1-t^2) {d\over dt} - {4(n+1)^2 \over 1-t^2} +   { (2+ m^2) (1+\xi) \over \xi  - t^2}  \right] \,   \tilde \chi_n \, = \, 0\ .
\eea
This is an  equation with  five  regular singular points, at $t= \pm 1$, $t= \pm\sqrt{\xi}$ and  $t = \infty$,
 all of which  lie outside the open interval $(-1, 1)$ where $\tilde\chi_n(t)$ should be defined.
The AdS$_5\times$S$^5$ equation is  recovered in the limit $\Delta\phi\to 0$, i.e. $\xi\to \infty$.
\vskip 2mm

 The number
  of singularities  of
  (\ref{Heun2}) can be reduced by exploiting its symmetry  under the reflection $t\to -t$.
   Thanks to this symmetry,
   the eigenfunctions can be chosen to be either even or odd functions of $t$.
 Changing variable to $s = t^2$ leads to
\bea\label{houn}
\bigg[ 4 s (1-s) \frac{d^2}{ds^2}   + 2 (1 - 3s) \frac{d}{ds} - {4(n+1)^2\over 1-s} +  { (2+m^2) (1+\xi) \over \xi -   s}  \bigg] \, \tilde \chi_n \
\ =\  0  \ ,
\eea
with even/odd solutions  distinguished by their behavior at $s\to 0$:
\bea\label{evenodd}
 \tilde\chi_n^{\rm even} = {\rm analytic}\ , \qquad  \tilde\chi_n^{\rm odd} =   ({\rm analytic})\, \times \sqrt{s}\ .
 \eea
 Equation (\ref{houn}) has   four regular singular points,  at $s=0, 1, \xi$ and $\infty$.  [That the point $s=\infty$
  is   regular   can be seen after a  change of variables  $s^\prime = 1/s$.]  Such  differential equations
   are  known in the mathematics literature as Heun's equations \cite{heun,localHeun}.
  They arise in a variety of physical contexts, for instance as  eigenvalue equations in the background  of
the Kerr-Newman-de Sitter black hole \cite{Suzuki:1999nn},  of large AdS black holes in five dimensions \cite{Musiri:2003rs},
 of the  toric Sasaki-Einstein manifolds $L^{a,b,c}$ \cite{Oota:2005mr}, or  in the study
 of  RG flows between 2D CFTs  \cite{Berg:2002hy} and of
 slowly-rotating homogeneous stars \cite{stars}
 (for more physics applications see the references in \cite{heun, Hortacsu:2011rr}). Heun's equation is  the
 next   in the series of canonical Fuchsian equations on the Riemann sphere, after the hypergeometric
 equation which has three regular singular points.
 \vskip 1mm

 Near a regular singular point, the two (local) solutions of a Fuchsian
 equation behave as  $(s-s_0)^{\nu_1} f_1(s)$ and $(s-s_0)^{\nu_2} f_2(s)$, where $\nu_1$ and $ \nu_2$
 are called the singularity exponents,  and $f_1, f_2$ are analytic at $s_0$.\footnote{Except when
  $(\nu_2-\nu_1)$ is a non-negative integer,
 in which case $f_1$  has a logarithmic singularity.}
  In  the case at hand, the exponents at the singular point $s_0=0$ are  $\nu_1=0$ and $\nu_2= 1/2$,
 corresponding respectively  to the even and the odd solutions.
At the singular point $s_0 = 1$,  on the other hand, the exponents  are
$\pm (n+1)$.  Since  the normalizability condition  (\ref{bc2})  requires that
  \bea
    \int_0^1 {ds\over \sqrt{s}}\, {\xi \over \xi - s}\, \vert \tilde\chi_n \vert^2 < \infty\ ,
  \eea
  solutions singular at $s_0=1$ must be  excluded. Thus, to determine the spectrum we need to  find
 the  values of $m^2$  for which  the even (odd) solutions of (\ref{houn})  are analytic at    $s=1$ .
 [Solutions that are analytic in a region encompassing two singular points,  are called Heun's functions in the
mathematics  literature.]

 \vskip 1mm

  For the generic Heun equation  it is not known how to solve  this  two-point connection problem
 in closed form.  Semi-analytic expressions for $\tilde\chi_n$ can be written down  as series of hypergeometric functions
 whose coefficients obey simple recurrence relations \cite{heun}. These series converge in a region around a given singular point,
 so  to determine $m^2$ in the case at hand one must  match the expansions around the points $s=0$ and $s=1$
  (for an example of this procedure see \cite{Suzuki:1999nn}). For the purposes of this work,
  simpler numerical algorithms will be sufficient. Before turning to the numerical solution, we first bring
  equation (\ref{houn}) to the canonical Heun form
     \begin{align}
\label{Heun}
\frac{d^2g}{d^2s}  + \bigg(\frac{\gamma}{s} +  \frac{\delta}{s-1} +  \frac{\epsilon}{s - \xi} \bigg) \frac{dg}{ds}
 + \frac{\alpha \beta s - q}{s (s-1)(s-\xi)} g = 0\ ,
\end{align}
where the  parameters must obey  the constraint $\gamma + \delta + \epsilon = \alpha + \beta + 1$.
Our equation is brought to this form  by the  wavefunction redefinition
  \bea
  \tilde\chi_n(s) \equiv (1-s)^{n+1} (\xi-s ) \, g (s)\ ,
  \eea
and with  the following  identification of the parameters:
  \begin{align}\label{paramident}
 \alpha = n+2  \, , \qquad  & \beta =  n+ \frac{5}{2} \, , \qquad
\gamma = \frac{1}{2} \, , \qquad \delta = 2n+ 3\,  , \qquad \epsilon = 2 \, ,   \no\\
4q &=  \xi (2n+1)(2n+4) - (1+\xi) m^2\  .
\end{align}
Note that the roles of $\alpha$ and $\beta$ in the canonical form of Heun's equation
can be swapped. Note also that since the exponents of $\tilde\chi_n$ at $s=1$ were
$\pm (n+1)$, those of $g$ at this singular point are $-2n-2$ and $0$.  The normalizable solutions have
 $g(s)$  analytic at  $s=1$,  and analytic or with a square-root singularity at $s=0$.



 \subsection{Numerical solution}
 \label{sec:numspec}

The Heun equation admits power series solutions around each of its singular points.
 We write the power-series solution around the point $s = 0$ as
\begin{align}
\label{pss}
g_1(n,m^2,\xi;s) \equiv   {Hl}\, (\alpha,\beta,\gamma,\delta;q;\xi;s)
 =
 \sum_{i=0}^\infty c_i s^i \qquad {\rm with} \qquad c_0 = 1 \, .
\end{align}
The  $c_i$ are then determined by the three-term recursion relation
\begin{align}
\label{recursion}
0 &= -q c_0 + \xi \gamma c_1 \ , \nonumber \\
0 &= P_i c_{i-1} - (Q_i + q) c_i + R_i c_{i+1}\ ,
\end{align}
where the coefficients in the lower line are given by
\begin{align}
P_i &= (i-1 + \alpha)(i-1 + \beta) \ ,  \nonumber \\
Q_i &= i[(i-1+\gamma)(1+\xi) + \xi \delta + \epsilon] \ ,  \nonumber \\
R_i &= (i+1)(i+\gamma) \xi \ ,
\end{align}
for $i=1,2 \cdots $.
The second, linearly-independent solution of Heun's equation (\ref{Heun}) can then
be written in terms of the same function $Hl$,
\bea
g_2(n,m^2,\xi;s) = s^{1-\gamma} Hl\, (\alpha^\prime,\beta^\prime ,2- \gamma,\delta;q^\prime ;\xi;s)
\eea
 with the following transformed arguments:
\bea
\alpha^\prime = \alpha-\gamma+1\ , \qquad \beta^\prime = \beta-\gamma+1\ ,
\qquad q^\prime = q + (1-\gamma)(\epsilon + \xi \delta)\ .
\eea
Recall that $n$ was a label for the Fourier modes in the $y$ direction, $\xi$ parametrizes the dilaton jump,
the $AdS_4$ mass is  $m^2$, and the remaining parameters are given in eq. (\ref{paramident}).
Note also that for $\gamma = 1/2$  the  two solutions correspond to the two possible behaviors
  (\ref{evenodd}), i.e. $g_1$ is the even and $g_2$ the odd solution. Our problem is to find the values of $m^2$ for
  which these solutions are regular at $s=1$.
\vskip 1mm

The asymptotic behavior near  $s = 1$ can be studied most easily by making a linear fractional  change of variable
\begin{align}
\hat s = \frac{\xi (s-1)}{s- \xi} \,,
\end{align}
and a redefinition of the  function $\hat g(\hat s) = (\hat s - \xi)^{-\alpha} g(\hat s)$.
This M\"obius transformation
maps Heun's equation into itself, swapping the singular points $0 \leftrightarrow 1$ and $\xi \leftrightarrow \infty$, and
transforming the  parameters as follows:
\begin{align}
&\hat \alpha = \alpha, \qquad \hat \beta = 1+\alpha-\epsilon, \qquad \hat \delta = \gamma, \qquad \hat \gamma = \delta, \no\\
& \hat \epsilon = 1+\alpha-\beta, \qquad \hat q = q - \alpha(\beta - \delta) \, .
\end{align}
The  solution of the original equation that is regular at $s=1$ can be written as
\bea
g_{\rm reg} (n,m^2,\xi;s) =  (\xi - \hat s)^{\hat \alpha} Hl(\hat \alpha, \hat \beta, \hat \delta, \hat \gamma, \hat \epsilon;\hat q;\xi; \hat s)\  ,
\eea
while the second solution  is singular and must be excluded.  The normalizable  wavefunctions are thus  the
 ones for which $g_1 = A_1 g_{\rm reg} $ or $g_2 = A_2 g_{\rm reg} $, where $A_1$ and $A_2$ are
 proportionality  constants.
  These conditions
 determine the allowed values of $m^2$ for given values of $n$ and $\xi$.
\vskip 1mm

 There are no general techniques which are currently known
 for the analytic solution of  this problem.\footnote{An approach along the lines of  reference \cite{Suzuki:1999nn} may,  however,  be
 worth trying.
 }
The following simple numerical algorithm will be, however, sufficient for our purposes here.
Using the recursion relation  (\ref{recursion}) we compute the series expansions
of $g_{1,2}$ and $g_{\rm reg}$,  around the singular points $0$ and $1$,  to a high order.  We then impose
the following matching condition for the even or odd solutions:
\begin{align}
\frac{\p}{\p s} \ln \frac{g_{1}}{g_{\rm reg}} \bigg|_{s_0} = 0 \qquad {\rm or}
 \qquad \frac{\p}{\p s} \ln \frac{g_{2}}{g_{\rm reg}} \bigg|_{s_0} = 0 \, ,
\end{align}
where $s_0$ is a midway point at which all series expansions are convergent.
The results for $n=0$ and different values of $\xi$ are shown in figure \ref{fig:massspect}.
As can be seen, all the masses decrease monotonically as $\xi$ increases, and they all approach zero
as $\xi$ approaches $1$. The behavior for $\xi$ close to $1$, corresponding to large $\Delta\phi$, is shown
 in figure \ref{fig:linearfit}.  As discussed in section 4.2, a nearly-flat fifth dimension grows to
  size $\sim \Delta\phi$ in this limit.  Thus the  spacing  between the masses of neighboring modes
  should be constant and of order  $ \Delta \phi^{-1}$, and this  is confirmed by the numerical analysis.

 \vskip 1mm

\begin{figure}[t]
\vspace{-0.5cm}
\centering
\includegraphics[width=0.47\textwidth]{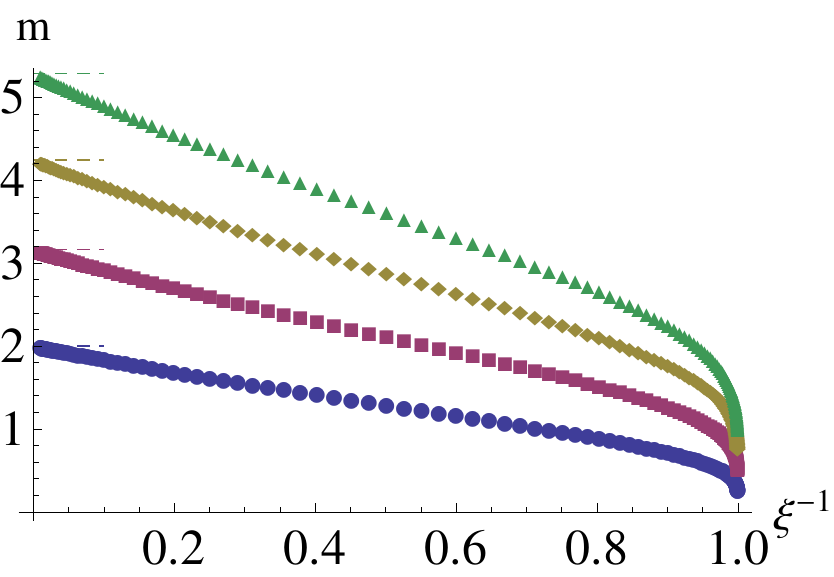}
\caption{\footnotesize
Numerical results for the AdS$_4$ masses of the first four normalizable modes for $n=0$.
 The dashed lines at $\xi\simeq \infty$ correspond to the $AdS_5$ values $m= 2, \sqrt{10}, \sqrt{18}, \sqrt{28}$.}
\label{fig:massspect}
\end{figure}

\begin{figure}[t]
\vspace{1cm}
\centering
\includegraphics[width=0.47\textwidth]{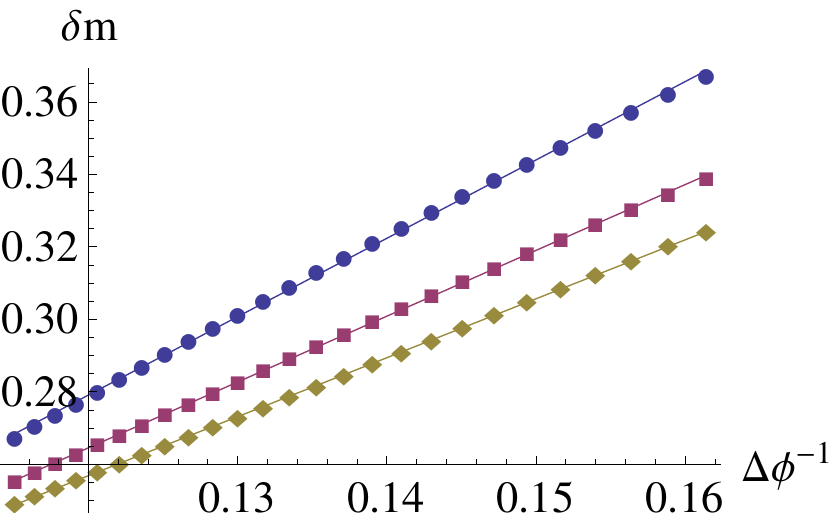}
\caption{\footnotesize
Numerical results for the mass spacing, $\delta m$,  as a function of $\Delta \phi^{-1}$
for large $\Delta\phi$.  Shown are the spacings between the first two modes (blue circles), between the third and the second mode (purple squares)
and between the fourth and third modes (yellow diamonds). The solid lines are a linear fit to the numerical results.}
\vspace{5mm}
\label{fig:linearfit}
\end{figure}

 \vskip 1mm

 This limiting situation can be analyzed more easily by going back  to
 the original form (\ref{Heun1})  of the eigenmode equation. Expanding this equation for large $\Delta\phi$
  gives
 \bea
 \left( -{1\over 2} {d^2\over dx^2}   + 2 e^{2\vert x\vert - \Delta\phi} \right) \tilde\chi_0\  \simeq \  m^2 \tilde\chi_0\ ,
 \eea
 where we have set $n=0$ and we assumed that $m\ll 1$. The solutions are plane waves,  $\tilde \chi_0 \simeq e^{\pm \sqrt{2m} x}$ ,
 cutoff by the exponentially-rising potential at $x\simeq \pm \Delta\phi/2$. The potential wall discretizes the low-lying mass spectrum
 in units of $\Delta\phi^{-1}$ as advertized.
 This completes our analysis of the Janus spectrum.



\section{Discussion}

    A lot of effort has gone,  in recent years,  into finding a string-theory realization of the
    so-called   model-I   of Randall and Sundrum \cite{Randall:1999ee}.
    A  popular embedding     \cite{Giddings:2001yu}, for example,     uses the
    Klebanov-Strassler throat  geometry  \cite{Klebanov:2000hb} capped-off by a compact
    Calabi-Yau manifold. Despite some notable differences,  the gravitational physics in
    this setting resembles that of ordinary Kaluza-Klein compactifications
     (see for instance refs.   \cite{Firouzjahi:2005qs, Noguchi:2005ws, Shiu:2007tn}).
    There exists, in particular, a normalizable zero mode and a finite mass gap for the higher  spin-2 excitations.
   \vskip 1mm

      The backgrounds considered in this work are qualitatively different,  since the internal space is not compact
      and the graviton has a non-vanishing  AdS$_4$ mass.  They resemble in this respect the thin-brane geometry of the
      Karch-Randall model \cite{KR1},   a prototypical example of  (quasi-)localized  gravity.
      A crucial aspect of this model is that the  excited modes  contribute to the long-range force, but this
      contribution is  suppressed by an exponentially-rising warp factor.
              The lack of a  compelling effective theory for massive gravity in four dimensions probably means that
       efforts to  embed  this  model  in string theory are doomed to fail.
      Nevertheless, given the importance of the question,  even failure can be instructive and all  avenues should be
      pursued. The   exact interface backgrounds
      of D'Hoker et al \cite{DEG1, DEG2} allow us to revisit this question in a controlled setting, and this
      has motivated the present work.
   \vskip 1mm

         One of our tasks in this paper was  to carefully set up  the spectral problem in the above  background
         geometries, in sections \ref{sec2} and \ref{sec5}.   The exercise is  straightforward,  but
          performing carefully  all steps
        allowed us  to tie up  loose ends,  and to emphasize the differences with the more familiar  cases of
        flat-brane warped geometries  and of  direct-product AdS$_4$ compactifications.
         In this paper  we have  solved this
        spectral problem only in the simplest case of  the supersymmetric Janus solution, for  which
          the eigenmode equation is separable.
           Our discussion is, however, also the starting point  for the study  of the more involved five-brane solutions,
            whose spectral problem is inherently 2-dimensional and thus requires more sophisticated numerical and
            analytic techniques \cite{tocome}.  Finally, although much of our discussion has been guided by the issue
            of gravity localization, the spectral problem for spin-2 modes is of  broader interest from the perspective
             of defect CFTs.
            One  question, to which we hope to return, is the precise relation between this spectrum
            and the energy-transfer properties of the holographically-dual  defect walls.
        \vskip 1mm

       \vskip 0.5cm

{\bf Aknowledgements}:
We have benefited from discussions with   Benjamin  Assel,  Iosif  Bena,  Cedric Deffayet,  Eric D'Hoker, Elias Kiritsis,
Joao Penedones,
Tassos Petkou, Giuseppe Policastro, Slava Rychkov, Paul Townsend,
 Jan Troost and  Ed  Witten.   J.E. is supported by the FWO - Vlaanderen, Project No. G.0235.05, and by
 the ``Federal Office for Scientific, Technical and Cultural Affairs through the Inter-University Attraction Poles
 Programme,"  Belgian Science Policy P6/11-P.


\appendix

\section{Summary of Type IIB supergravity}
\label{Summary of Type IIB supergravity}
\renewcommand{\theequation}{A.\arabic{equation}}
\setcounter{equation}{0}

We briefly review here the type-IIB supergravity  Bianchi identities
and field equations, as well as the supersymmetry variations,   for vanishing
fermionic  fields. Our  conventions are those used in \cite{DEG1,DEG2},
 except for a rescaling of the 4-form potential by a factor $4$ so as to agree
 with the standard conventions.  The bosonic fields are  the metric $g_{MN}$,
the complex axion-dilaton scalar $B$,   the complex 2-form $B_{(2)}$
and the real 4-form $C_{(4)}$.
To simplify the form of the field equations and Bianchi identities
we introduce the following composite fields:
\bea
\label{sugra1}
P  = f^2 d B\ , \qquad Q  =  f^2\,  {\rm Im} ( B d  \bar B)\ \qquad {\rm where}\ \
  f^2=(1-|B|^2)^{-1}
 \ ,
\eea
and the complex 3-form and real 5-form field strengths
\bea
\label{GF5}
G & = &  f(F_{(3)} - B \bar F_{(3)} )\ ,
\no \\
F_{(5)} & = & dC_{(4)} + { i \over 4} \left ( B_{(2)} \wedge \bar F_{(3)}
- \bar B_{(2)} \wedge  F_{(3)} \right )\ ,
\eea
where $F_{(3)} = d B_{(2)}$.
The scalar field $B$ is related to the complex string coupling $\tau$,
the axion $\chi$, and the dilaton $\phi$   by
\bea
\label{Btau}
B = {1 +i \tau \over 1 - i \tau }\ ,  \hskip 1in \tau =  \chi + i e^{- 2\phi}\ .
\eea
In terms of the composite fields $P,Q$, and $G$, the  Bianchi identities
read
\bea
0 &=& dP-2i Q\wedge P\ , \no
  \\
0 &=& d Q + i P\wedge \bar P\ ,
\no  \\
0 &=& d G - i Q\wedge G +  P\wedge \bar G\ ,
\no  \\
0 &=& d F_{(5)} -  {i\over 2} G \wedge \bar G\ .
\label{bianchi4}
\eea
The field strength $F_{(5)}$ is furthermore required to be self-dual,
\bea
\label{SDeq}
 F_{(5) } = * F_{(5) }\ .
\eea
The field equations are given by
\bea
0 & = &
\nabla^\mu  P_\mu  - 2i Q^\mu  P_\mu
+ {1\over 24} G_{\mu \nu \rho  }G^{\mu \nu \rho }\ ,
\no
\\
0 & = &
\nabla ^\rho   G_{\mu \nu \rho  } -i Q^\rho  G_{\mu \nu \rho }
- P^\rho  \bar G_{\mu \nu \rho  }
+ {2\over 12} i F_{(5)\mu \nu \rho \sigma \kappa  }G^{\rho \sigma \kappa }\ ,
\no
\\
0 & = & R_{\mu \nu  }
- P_\mu   \bar P_\nu   - \bar P_\mu   P_\nu
- {1\over 96} F_{(5) \mu  \rho \sigma \kappa \gamma} F_{(5)\nu }{}^{\rho \sigma \kappa \gamma}
\no \\ && \hskip .5in
- {1\over 8} (G_\mu  {} ^{\rho \sigma  } \bar G_{\nu  \rho \sigma  }
+ {\bar G_\mu } {} ^{ \rho \sigma  } G_{\nu  \rho \sigma  })
+{1\over 48 } g_{\mu \nu  } G^{\rho \sigma \kappa  } \bar G_{\rho \sigma \kappa  }
\label{Eeq}
\eea

 \vskip 2mm

Type-IIB supergravity is invariant under $SU(1,1) \sim SL(2,{\bf R})$ transformations
which leave  $g_{\mu \nu}$ and $C_{(4)}$ invariant, act  as  M\"obius
transformations on the field $B$, and transform linearly the 2-form  $B_{(2)}$,
\bea
\label{stransf}
B           & \to & B^s = {u B  + v \over \bar v B + \bar u}\ ,
\no \\
B_{(2)} & \to & B_{(2)}^s = u B_{(2)} + v \bar B_{(2)}\ ,
\eea
with $u,v \in {\bf C}$ and $\bar u u - \bar v v=1$. In this non-linear realization of
$SU(1,1)$, the field $B$ takes values in the coset  $SU(1,1) /U(1)_q$,
and the fermions $\lambda$ and $\psi_\mu$ transform linearly under
the isotropy gauge group $U(1)_q$ with composite gauge field $Q$. The transformation
rules for the composite fields are,
\bea
\label{su11a}
P & \to & P^s = e^{2 i \theta} P\ ,
\no \\
Q & \to & Q^s = Q + d \theta\ ,
\no \\
G & \to & G^s = e^{i \theta} G\ ,
\eea
where the phase $\theta$ is defined by
\bea
\label{su11b}
e^{i \theta} = \left ( {v \bar B + u \over \bar v B + \bar u} \right )^\half\ .
\eea
Written like this the transformation rules clearly exhibit the $U(1)_q$ gauge
variation  that accompanies the  global $SU(1,1)$ transformations.


\end{document}